\documentstyle[11pt]{article}
\input amssymb.sty

\newcounter{mo}

\newcounter{bk}

\newcommand{\tr}{{\rm tr}}
\newcommand{\ad}{{\rm ad}}

\newcommand{\ti}[1]{\tilde{#1}}
\newcommand{\om}{\omega}

\newcommand{\al}{\alpha}
\newcommand{\te}{\theta}
\newcommand{\vth}{\vartheta}
\newcommand{\be}{\beta}
\newcommand{\la}{\lambda}
\newcommand{\La}{\Lambda}

\newcommand{\ve}{\varepsilon}
\newcommand{\ep}{\epsilon}
\newcommand{\vf}{\varphi}

\newcommand{\ga}{\gamma}
\newcommand{\ze}{\zeta}
\newcommand{\si}{\sigma}

\def\bfe{{\bf e}}
\def\bff{{\bf f}}

\def\bfS{{\bf S}}
\def\bfJ{{\bf J}}
\def\bfC{{\bf C}}

\def\hS{\hat{S}}
\def\bhS{\bf\hat{S}}
\def\cO{{\cal O}}

\def\cA{{\cal A}}
\def\cP{{\cal P}}
\def\cR{{\cal R}}
\def\hP{\hat{P}}

\def\mC{{\mathbb C}}
\def\mZ{{\mathbb Z}}
\def\mR{{\mathbb R}}
\def\mN{{\mathbb N}}

\newcommand{\beq}[1]{\begin{equation}\label{#1}}
\newcommand{\eq}{\end{equation}}
\newcommand{\beqn}[1]{\begin{eqnarray}\label{#1}}
\newcommand{\eqn}{\end{eqnarray}}
\newcommand{\p}{\partial}
\newcommand{\di}{{\rm diag}}
\newcommand{\oh}{\frac{1}{2}}

\newcommand{\GLN}{{\rm GL}(N,{\mathbb C})}
\def\sln{{\rm sl}(N, {\mathbb C})}

\newcommand{\gln}{{\rm gl}(N, {\mathbb C})}

\def\f1#1{\frac{1}{#1}}

\newcommand{\bp}{\bar{\partial}}

\def\frak{\mathfrak}
\def\gg{{\frak g}}

\def\gS{{\frak S}}
\def\gA{{\frak A}}

\newcommand{\ran}{\rangle}
\newcommand{\lan}{\langle}

\renewcommand{\theequation}{\thesection.\arabic{equation}}
\newtheorem{predl}{Proposition}[section]

\newtheorem{rem}{Remark}[section]
\newtheorem{cor}{Corollary}[section]
\newtheorem{lem}{Lemma}[section]
\newtheorem{theor}{Theorem}[section]

\newcommand{\AmS}{{\protect\the\textfont2
  A\kern-.1667em\lower.5ex\hbox{M}\kern-.125emS}}

\begin{document}

\vspace{0.3in}
\begin{flushright}
 ITEP-TH-30/03\\
{\large{August 22, 2003}}\\
\end{flushright}
\vspace{10mm}
\begin{center}
{\Large{\bf
Bihamiltonian structures and quadratic algebras in hydrodynamics and
on non-commutative torus}
}\\
\vspace{5mm}
B.Khesin
\\
{\sf Max Planck Institute of Mathematics, Bonn, Germany}\\
{\sf University of Toronto, Toronto, Canada}\\
{\em e-mail: khesin@math.toronto.edu}
\\
A.Levin
\\
{\sf Max Planck Institute of Mathematics, Bonn, Germany}\\
{\sf Institute of Oceanology, Moscow, Russia}\\
{\em e-mail: alevin@wave.sio.rssi.ru}
\\
M. Olshanetsky
\\
{\sf Max Planck Institute of Mathematics, Bonn, Germany}
\\
{\sf Institute of Theoretical and Experimental Physics, Moscow, Russia,}\\
{\em e-mail: olshanet@gate.itep.ru}\\

\vspace{5mm}
\end{center}

\begin{abstract}
 We demonstrate the common  bihamiltonian nature of several
integrable systems. The first one is an elliptic rotator
that is an integrable Euler-Arnold top
on the complex group $\GLN$ for any $N$, whose inertia ellipsiod
is related to a choice of an elliptic curve. Its bihamiltonian structure
is provided by the compatible linear and quadratic Poisson brackets,
both of which are governed by the Belavin-Drinfeld classical
elliptic $r$-matrix. We also generalize this bihamiltonian construction of
integrable Euler-Arnold tops to several infinite-dimensional groups,
appearing as  certain large $N$ limits of $\GLN$.
These  are the group of a non-commutative torus (NCT) and the group of
symplectomorphisms $SDiff(T^2)$ of the two-dimensional torus.
The elliptic rotator on symplectomorphisms gives an elliptic version
of an ideal 2D hydrodynamics, which turns out to be an integrable system.
In particular, we define the quadratic Poisson algebra on the space of
Hamiltonians on $T^2$ depending on two irrational numbers.
In conclusion, we quantize the infinite-dimensional quadratic
Poisson algebra in a fashion similar to the corresponding
finite-dimensional case.
\end{abstract}

\tableofcontents

\section{Introduction}

One of the most interesting problems in the theory of integrable systems is
a description of infinite-dimensional integrable systems with two or more
space variables. In this paper we consider integrable hierarchies in $(2+1)$
dimension, which we call {\sl elliptic} and {\sl modified hydrodynamics}.
These are close relatives
of the ideal hydrodynamics on a two-dimensional torus, which is known to be
a highly non-integrable system. For instance, even the 4-vortex
approximation of the ideal fluid dynamics is non-integrable \cite{Z}.
We obtain these integrable versions of hydrodynamics 
starting from finite-dimensional integrable systems whose
dimension of the phase space goes to infinity. Our approach is somewhat 
similar to the derivation of Toda field theory from the open Toda chain. 
In contrast with the Toda theory, however, we obtain non-local systems.

The classical Euler (or, rather, Helmholtz) equation for the motion of
an ideal fluid on the standard two-torus is
$$
\partial_t\bfS=\{\bfS, \Delta^{-1}\bfS\}\,,
$$
where $\bfS$ is the vorticity function of the fluid flow,
$\{~\}$ is the Poisson bracket, and $\Delta$ is the Laplace operator
on $T^2$. The modified and elliptic hydrodynamics are defined on an
elliptic curve, i.e., on a torus with a complex structure fixed.
The corresponding equations are, respectively,
$$
\partial_t\bfS=\{\bfS, {\bar\partial}^{-2}\bfS\}
$$
and
$$
\partial_t\bfS=\{\bfS, \wp(\bp)\bfS\}\,,
$$
where $\bp$ is the corresponding operator of the complex structure,
and $\wp$ is the Weierstrass $\wp$-function. (We postpone the precise
description of $\wp(\bp)$ till Section \ref{DIFF-ER}.)
Note that to define the Laplace operator one needs to choose a metric,
while for the operator $\bp$ in the modified hydrodynamics is
defined by a complex structure on $T^2$.

Such a modification of the fluid inertia operator from the Laplace
operator $\Delta=\partial\bp$, which depends on a metric on $T^2$, to
$\bp^{2}$ or $\wp(\bp)^{-1}$, both of which depend on a complex structure 
on $T^2$, brings in the  integrability
and even the bihamiltonian structure for the systems.
We construct an infinite set of involutive integrals
of motion with respect to two  Poisson brackets.
One of the brackets is the standard
linear Lie-Poisson brackets on the dual space to the algebra of the
divergence-free vector fields on $T^2$. The other Poisson structure is a
quadratic Poisson algebra on  Hamiltonians of vector fields.
These two brackets are compatible and governed by the same classical
$r$-matrix. Furthermore, we describe a recursion procedure
for constructing the sequence of Hamiltonians for this linear-quadratic
bihamiltonian structure of the hierarchy of the elliptic hydrodynamics,
which thereby exhibits ``the strongest form of integrability.''

We come to this construction through the non-commutative deformation
of $T^2$ to the non-commutative torus (NCT).
The non-commutative deformation of the Lie algebra of vector fields is
isomorphic to the Lie algebra of NCT. The latter is a special
large $N$ limit of $\gln$. We start with the elliptic rotators on
$\GLN$ \cite{STSR,LOZ} and develop their bihamiltonian structure
based on the Belavin-Drinfeld classical $r$-matrix \cite{BD}.
The corresponding quadratic Poisson
algebra is the classical limit of the Feigin-Odesski algebra \cite{FO}.
We describe natural extensions of the elliptic rotators to the
infinite-dimensional groups, which preserve their main properties.
This brings us to bihamiltonian systems and integrable hierarchies
on the NCT, which are of interest by themselves.

As a byproduct, we also describe a quantization of the quadratic Poisson
algebra on NCT. In this way we obtain an infinite-dimensional
associative algebra
with quadratic relations depending on the complex structure on $T^2$
and the Planck constant $\hbar\in T^2$.

Finally, the commutative ("dispersionless") limit of the
elliptic rotators on NCT leads to the
desirable bihamiltonian hierarchies of the elliptic and modified
hydrodynamics.


\section{Main results}
\setcounter{equation}{0}

{\sl 1. The general setup \cite{Ar1, Ar}}\\
Let $G$ be a Lie group and $\gg$ its Lie algebra.
Consider an invertible linear operator $\bfJ$ that maps
the coalgebra $\gg^*$ to $\gg$.
Its inverse operator  $\bfJ^{-1}$ is called {\sl the inertia tensor}.
The Euler-Arnold top corresponding to the group $G$ is the
Hamiltonian system on $\gg^*$ with respect to
the linear Lie-Poisson brackets on $\gg^*$ and the Hamiltonian function
given by the quadratic form
$$
H=-\oh\lan \bfS,\bfJ(\bfS)\ran\,,~~\bfS\in\gg^*\,,
$$
where $\lan~,~\ran$ stands for the pairing between $\gg$ and $\gg^*$.
 Namely, the variation $\nabla H$  can be regarded as
an element of $\gg$ and the corresponding Hamiltonian
 equation of motion is as follows:
$$
\p_t\bfS=\{H,\bfS\}:={\rm ad}^*_{\nabla H}\bfS\,.
$$
Recall  that the Lie-Poisson
brackets are degenerate on $\gg^*$ and their symplectic leaves are
coadjoint orbits of $G$. To descend to a particular coadjoint orbit
$\cO$ one should fix the values of Casimirs for the linear bracket.

For some special choices of $\bfJ$ the system
becomes completely integrable. Some of the examples are Manakov's tops on
$SO(N)$ \cite{Man}, their limit $N\to\infty$ found by Ward \cite{Wa},
 the Korteweg--de Vries equation
\cite{Kh, S}, as well as the Camassa--Holm and Hunter--Saxton equations
on the Virasoro group \cite{KM}.
(In the infinite-dimensional case, the invertibility of $\bfJ$
is understood as  Ker$\bfJ=0$.)

\bigskip

\noindent
{\sl 2.  Elliptic Rotators on $\GLN$}\\
 The elliptic
rotator (ER) on $\GLN$ is an integrable Euler-Arnold top on this group,
whose inertia operator is constructed with the help of the
Weierstrass $\wp$-function of an auxilliary elliptic curve, see
Ref.\,\cite{STSR,LOZ,Be}. Namely, let $T_\al$, $\al=(\al_1,\al_2)$ be
the basis of $\sln$
 (\ref{AA3},) $\al_j=(0,1,\ldots,N-1),\,\al\neq(0,0)$.
 The structure constants in this basis are
$$
\bfC_\te(\al,\be)=\f1{\pi\te}\sin\pi\te(\al\times\be)\,,~~~\te=k/N\,,
$$
where $1\leq k<N$ and $k,N$ are coprime, see (\ref{AA4}).
 For $\bfS=\sum_\al S_{-\al}T_\al\in\sln^*$
the linear Poisson brackets assume the form
\beq{LB}
\{S_\al,S_\be\}=\bfC_\te(\al,\be)S_{\al+\be}\,.
\eq
Let $\wp(x;\tau)$ be the Weierstrass
function on the elliptic curve $E_\tau=\mC/(\mZ+\tau\mZ)$.
 Consider its values  $\wp_\te(\al)=\wp((\al_1+\al_2\tau)\te;\tau)$
on the lattice parametrized by $\alpha$, the labels of the
$\sln$-basis.
The inverse inertia operator is defined as
\beq{bfJ}
\bfJ~:~S_\al\to J_\al S_\al\,,~~J_\al=\wp_\te(\al)\,,
\eq
It was proved in Ref\,.\cite{STSR} that the equations of motion
\beq{0}
\p_tS_\al=\sum_\ga S_{\al-\ga}\wp_\te(\ga)S_\ga\,,
\eq
defined by the Hamiltonian $H=-\oh\lan \bfS,\bfJ(\bfS)\ran
=2\pi^2\te^2\tr(\bfS\cdot\bfJ(\bfS))$ and the brackets (\ref{LB}),
have the Lax representation with
the Lax operator $L^{rot}(\bfS,z)\in\sln$ depending on the spectral
parameter
$z\in E_\tau$. The traces $\tr(L^{rot}(\bfS,z))^k$, $k=2,\ldots,N$ being
expanded in the basis of elliptic functions on $E_\tau$ produce the
involutive integrals of motion and the hamiltonian $H$ is among them.

We call the commuting flows defined by these integrals {\sl the ER hierarchy}.

Let $r(z),\,z\in E_\tau$ be the Belavin-Drinfeld classical elliptic r-matrix,
see \cite{Be,BD} and Section \ref{GLN-ER} below.
The main result, which we will also generalize to the infinite-dimensional
situation, can be formulated as follows.
\begin{theor}\label{main-th}
{\bf i)} In terms of the $r$-matrix, the linear brackets (\ref{LB}) can be
written in the form
\beq{1}
\{L_1^{rot}(z),L_2^{rot}(w)\}_1=[r(z-w),L_1^{rot}(z)+L_2^{rot}(w)]\,.
\eq
{\bf ii)} Consider the  phase space extended by a new variable $S_0$, which has
zero bracket with all the rest. The same $r$-matrix defines
the quadratic Poisson algebra $P_{N,\te,\tau}$ related
to $\GLN$ (\ref{1.1}), (\ref{2.1}):
\beq{2}
\{L_1(z),L_2(w)\}_2
=[r(z-w),L_1(z)\otimes L_2(w)]\,,~~{\rm where }~~L(z)=S_0 Id+L^{rot}(z)\,.
\eq
{\bf iii)} The above two brackets are compatible, i.e.
any linear combination of them is a Poisson bracket.\\
{\bf iv)}
There exists a sequence of  integrals of motion in involution with
respect to each of the two brackets $\{h_j,h_k\}_{1,2}=0$
(here the lower indices refer to the linear and the quadratic brackets
respectively). They provide {\sl the bihamiltonian structure} of the
elliptic rotator (ER) hierarchy
$$
\{h_{j+1},\bfS\}_2=-\{h_j,\bfS\}_1\,.
$$
\end{theor}
The first two statements of Theorem for $\GLN$ are well known \cite{FO,Skl}.
Apparently, the bihamiltonian structure of the GL$_N$ elliptic rotators
is new, and it gives the following

\begin{cor}
The Casimirs with respect to one of the brackets
generate non-trivial dynamics with respect to the other.
\end{cor}
 In particular, the functional
 $S_0$, being the Casimir element of the linear
brackets (\ref{LB}), leads to the equations
\beq{4}
\p_t\bfS=\{S_0,\bfS\}_2\,,
\eq
that coincide with (\ref{0}).
\smallskip

It turns out that Theorem \ref{main-th} also holds in the
 infinite-dimensional situation presented below.

\bigskip

\noindent
{\sl 3.  Elliptic Rotators on the Non-commutative Torus}\\
Consider the $sin$-algebra $sin_\te$, called also the Lie algebra of the
non-commutative torus (NCT).
The Poisson brackets on the Lie coalgebra $sin^*_\te$ of the NCT has the
form (\ref{LB}) with irrational number $0\leq \te<1$ and the basis
$T_\al=const\cdot\exp(2\pi i\al_1x_1)*\exp(2\pi i\al_2x_2)$
is parameterized by the infinite lattice $\al\in\mZ\oplus\mZ$.
Here $\exp(2\pi ix_1),\exp(2\pi ix_2)$ are the generators of the NCT and $*$
is the Moyal multiplication (see Appendix C).
On the NCT we introduce the complex structure  $\bp_{\ep,\tau}$ depending on
$\tau$ (Im $\tau>0$), and two real numbers $\ep=(\ep_1,\ep_2),$
such that $\te\ep_a$ are
irrational  and $0<\te\ep_a\leq 1$ (see (\ref{cs})).
The inverse inertia operator $\bfJ$ is the pseudo-differential operator
$$
\bfJ(\bfS)(x)=\wp(\te\bp_{\ep,\tau})\bfS(x)\,,
$$
 where now $\bfS(x)=\sum_\al S_{-\al}T_\al(x)$.
The equation of motion in the form of the Moyal brackets has the
 form
\beq{3}
\p_t\bfS=\{\bfS,\wp(\te\bp_{\ep,\tau})\bfS\}^\te\,,
\eq
where $\{~\}^\te$ is the Lie--Poisson bracket on  $sin^*_\te$.
In Proposition 3.1 we find the Lax form of (\ref{3}).
We construct the classical $r$-matrix (\ref{1.20}) and prove the
 counterpart of the Theorem.
The quadratic Poisson algebra $P_{\te,\ep,\tau}$ (\ref{1.10}), (\ref{2.10})
on the NCT gives rise to the infinite bihamiltonian hierarchy, where the
equations (\ref{3}) correspond to the quadratic Hamiltonian  functional with
respect to the linear brackets. At the same time the equations (\ref{3})
can be interpreted as the Hamiltonian equations (\ref{4}) in the quadratic
Poisson algebra $P_{\te,\ep,\tau}$ with a linear Hamiltonian. For SL$(2)$ this
representation
was found in \cite{O} and for general case in \cite{BDOZ}.

\bigskip

\noindent
{\sl 4.  Elliptic Rotators on $SDiff(T^2)$}\\
In the limit $\te\to 0$ the Lie algebra of the NCT becomes
isomorphic to the Poisson algebra $\cA$ of smooth functions
on $T^2$ modulo constants $Ham(T^2)\sim C^\infty(T^2)/\mC$.
The algebra of Hamiltonians can be also described by the corresponding
Hamiltonian (or, divergence-free) vector fields on the torus.
More precisely,  the Lie algebra $SVect(T^2)$ of divergence-free vector fields
on $T^2$ is the (universal) cocentral extension of $Ham(T^2)$,
defined by the exact sequence
$$
0\to Ham(T^2)\to SVect(T^2)\to\mC^2 \to 0\,,
$$
where the image of $SVect(T^2)$ in $\mC^2$ is generated by the two fluxes
$\ep_1\p_1$, $\ep_2\p_2$.
For $\psi\in Ham(T^2)$ we have
$$
V_1(\psi)=-\f1{4\pi^2}\p_2\psi\,,~~V_2(\psi)=\f1{4\pi^2}\p_1\psi\,,
$$
and $V_{\{\psi,\psi'\}}
=[V(\psi),V(\psi)']$.
We construct the elliptic rotator on the Lie group
$SDiff(T^2)$ of area-preserving diffeomorphisms, corresponding to the
Lie algebra  $SVect(T^2)$.

Consider the dual space of linear functionals $Ham(T^2)^*$ in
the Fourier basis
$$
\{\bfe(\al\cdot x):=\exp 2\pi i(\al\cdot x)~|~(x=(x_1,x_2)\,,
~(\al\cdot x)=\al_1x_1+\al_2x_2\,,~\al_j\in\mZ\}\,,
$$
$$
\bfS=\sum_\al S_\al\bfe(-(\al\cdot x))\in Ham(T^2)^*\,.
$$
The Poisson structure on $\cA^*$ assumes the form (cf. (\ref{LB}))
$$
\{S_\al,S_\be\}_1=(\al\times \be)S_{\al+\be}\,.
$$
This Poisson structure is degenerate and has an infinite set of Casimirs
$$
C_k=\int_{T^2}\bfS^k\,, {\rm where }~~ k=2,3,\ldots\,.
$$
On a coadjoint orbit
$\cO\subset Ham(T^2)^*$ of $SDiff(T^2)$ the brackets are non-degenerate and
values of the Casimirs are fixed.

 Define the operator  $\bfJ$   (after the rescaling) as
the pseudo-differential operator $\wp(\bp_{\ep,\tau})$,
such that the Hamiltonian has the form
\beq{Hm}
H=-\oh\int_{T^2}\bfS\cdot\wp(\bp_{\ep,\tau})\bfS\,.
\eq
We prove that the corresponding Hamiltonian system admits
an infinite set of commuting integrals, see Section \ref{DIFF-ER}.

Recall that these elliptic rotators are parameterized, in particular, by the
auxiliary elliptic curve. Consider the simplest version of the ER,
corresponding to the rational degeneration of the elliptic
curve $E_\tau$. In the limit the (inverse) inertia operator becomes
$\bfJ=\bp_{\tau}^{-2}$ and the limiting Hamiltonian (\ref{Hm}) takes the form
$$
H=-\oh\int_{T^2}\bfS\cdot\bp_{\tau}^{-2}\bfS\,,
$$
see (\ref{qh}) and \cite{O}.

It is interesting to compare this with
the Euler equation for an ideal fluid on a torus, where
the inertia operator is the Laplacian (i.e.,  $\bfJ=\Delta^{-1}
=(\bp\partial)^{-1}$), and the Hamiltonian is
$$
H=-\oh\int_{T^2}\bfS\cdot\Delta^{-1}\bfS\,,
$$
where $\bfS$ plays the role of the vorticity \cite{Ar}.
While the Euler hydrodynamics equation is highly non-integrable, the
modification of the (inverse) inertia operator from $\bp\partial$
to $\bp_{\tau}^2$ leads to an integrable hierarchy.
We call the rational limit  {\sl the modified hydrodynamics} on $T^2$.
(In a sense, these systems are similar in spirit to the Etingof--Frenkel
current algebras \cite{EF} on $T^2$, which make use of the complex
structure on the elliptic curve to construct a central extension of
smooth currents on the torus.)

Finally, we prove the Theorem \ref{main-th} for the generic
systems of {\sl the elliptic hydrodynamics} (\ref{Hm}).
In particular, the quadratic Poisson algebra
$\cP_{0,\tau}$ (\ref{6.10}), (\ref{7.10})    on
$T^2$, along with the linear Poisson bracket,  provide
the bihamiltonian structure of the hierarchies of the elliptic and
modified hydrodynamics.


\section{$\GLN$-Elliptic Rotators}\label{GLN-ER}
\setcounter{equation}{0}

\subsection{Lax representation and integrals of motion}

{\sl 1. The system description}\\
The elliptic  $\GLN$-rotator
is defined on  the dual spaces of $\gln$ and $\sln$. Let
$\bfS=\sum_\al S_{-\al} T_\al$, where $T_\al$ is the
basis of $\sln$, see (\ref{AA3}). Then the Poisson structure on the dual space
$\sln^*$
is given by the linear Lie-Poisson brackets (\ref{AA4})
\beq{lb}
\{S_\al,S_\be\}_1=\bfC_\te(\al,\be)S_{\al+\be}\,,
\eq
where $\bfC_\te(\al,\be)$ are the structure constants of $\sln$ (\ref{AA4}).
 The Hamiltonian has the form
\beq{8.5}
H^{rot}=2\pi^2\te^2\tr(\bfS\cdot \bfJ(\bfS))
\equiv -\oh\sum_\ga S_\ga\wp_\ga S_{-\ga}\,.
\eq
It defines the equations of motion
\beq{em}
\p_t\bfS=[\bfJ(\bfS),\bfS].
\eq

 Let
$\bfJ~:~S_\al\to J_\al S_\al\,,$ be the (inverse) inertia operator with
$J_\al=\wp_\te(\al),$
where $\wp_\te(\al)$ are the values of the Weierstrass $\wp$-function
and defined by  (\ref{AA5}).  Then (\ref{em}) assumes
the form
\beq{8.5a}
\p_tS_\al=\sum_\ga S_{\al-\ga}S_\ga\wp_\te(\ga)\bfC_\te(\ga,\al)\,.
\eq

\bigskip
\noindent
{\sl 2. The linear brackets and $r$-matrix}\\
\begin{predl}\label{pr3.1} (cf.\cite{STSR})
The equations of motion (\ref{8.5}) have the Lax form
\beq{La}
\p_tL^{rot}(z)=[L^{rot}(z),M(z)]\,,
\eq
where
\beq{8.7}
L^{rot}=-\sum_\al S_\al\vf_\te(\al,z)T_\al\,,
\eq
\beq{8.7a}
M=-\sum_\al S_\al f_\te(\al,z)T_\al\,,
\eq
and the functions $\vf_\te$, $f_\te$ are given by (\ref{vf}), (\ref{f}).
\end{predl}
{\sl Proof.}\\
Substituting (\ref{8.7}) and (\ref{8.7a}) in the Lax equation  we obtain
$$
\p_tS_\al\vf_\te(\al,z)
=\sum_\ga S_{\al-\ga}S_\ga(\vf_\te(\al-\ga,z) f_\te(\ga,z)
-\vf_\te(\ga,z) f_\te(\al-\ga,z))\,.
$$
Now, using the explicit expressions  (\ref{vf}) and  (\ref{f})
for $\vf_\te$ and $ f_\te$ respectively, as well as
the Calogero functional equation (\ref{ad2}), we come to
$$
\p_tS_\al=\sum_\ga S_{\al-\ga}S_\ga\vf_\te(\al,z)
(\wp_\te(\al-\ga)-\wp_\te(\ga)),
$$
which coincides with (\ref{8.5a}). $\Box$

\bigskip
The Lie-Poisson brackets (\ref{lb}) admit the following r-matrix
description \cite{BD,Be,Skl,STS}.
Define the classical $r$-matrix by
\beq{6.1}
r_{N,\te,\tau}(z-w)=\sum_\ga\vf_\te(\ga,z-w)T_\ga\otimes T_{-\ga}\,,
\eq
where $\vf_\te$ is defined by (\ref{vf}).

\begin{lem}\label{le3.1}
\cite{Skl}.
The $r$-matrix (\ref{6.1}) satisfies the classical
Yang-Baxter equation
$$
[r_{N,\te,\tau}(z-w),r_{N,\te,\tau}(z)]+
[r_{N,\te,\tau}(z-w),r_{N,\te,\tau}(w)]
$$
\beq{YB}
+[r_{N,\te,\tau}(z),r_{N,\te,\tau}(w)]=0\,.
\eq
\end{lem}
{\sl Proof.}\\
We reduce the YB equation (\ref{YB})
 to the following functional equation
$$
\vf_\te(\ga,z-w)\vf_\te(\al,z)-\vf_\te(\al+\ga,z-w)\vf_\te(\al,w)
+\vf_\te(\al+\ga,z)\vf_\te(-\ga,w)=0\,,
$$
by using the commutation relations in $\sln$ (\ref{AA4b}).
It easy to see that it can be rewritten for the functions
$\phi_\te$ (\ref{phi}), since
all three terms have the common exponent $\bfe_\te((\ga_2+\al_2)z-\ga_2w)$.
Now, it coincides with the Fay identity (\ref{ad3}), where
we put
$$
u_1=(\ga+\al)\te\,,~~u_2=-\ga\te\,,~~z_1=z\,,~~z_2=w\,.
$$
$\Box$
\begin{predl}\label{pr3.2}
 In terms of the Lax operator (\ref{8.7a})
the brackets (\ref{lb}) are equivalent to the
following relation for the Lax operator
\beq{LYB}
\{L_1^{rot}(z),L_2^{rot}(w)\}_1=[r_{N,\te,\tau}(z-w),L^{rot}(z)\otimes Id+
Id\otimes L^{rot}(w)]\,,
\eq
 $$
L_1=L\otimes Id\,,~~
L_2=Id\otimes L\,.
$$
\end{predl}
{\sl Proof.}\\
To prove (\ref{LYB}) we rewrite it
 in the basis $T_\al\otimes T_\be$:
$$
\{S_{-\al},S_{-\be}\}\vf_\te(\al,z)\vf_\te(\be,w)
$$
$$
=S_{-\al-\be}C_\te(\al,\be)\left(
\vf_\te(-\be,z-w)\vf_\te(\al+\be,z)-
\vf_\te(\al,z-w)\vf_\te(\al+\be,w)
\right)\,.
$$
The same Fay identity (\ref{ad3}) implies that
$$
\vf_\te(\al,z)\vf_\te(\be,w)=
\vf_\te(-\be,z-w)\vf_\te(\al+\be,z)-
\vf_\te(\al,z-w)\vf_\te(\al+\be,w)\,.
$$
Thus we come to (\ref{lb}). $\Box$

\bigskip
\noindent
{\sl 3. The hierarchy of the Lax equations}\\
The Lax operator $L^{rot}$ (\ref{8.7}) has the following properties:\\
i)  $L^{rot}$ is an $\sln$-valued meromorphic function on $E_\tau$ with
a simple pole at the origin satisfying
$$
Res~ L^{rot}(z)|_{z=0}=\sum_\al S_{-\al}T_\al\,.
$$
ii) $L^{rot}$ satisfies the quasi-periodicity conditions
$$
L^{rot}(z+1)=QL^{rot}(z)Q^{-1}\,,~~
L^{rot}(z+\tau)=\La L^{rot}(z)\La^{-1}\,.
$$
These properties imply that
 $Tr(L(z))^k$ are doubly periodic functions with the poles up to
the order $k$. Thereby, they can be expanded
in the basis of the Weierstrass
function and its derivatives
\beq{8.21}
\tr(L^{rot}(z))^k=I_{0,k}+I_{2,k}\wp(z)+\ldots+I_{k,k}\wp^{(k-2)}(z)\,.
\eq
In particular, in this way we obtain the Hamiltonian
$$
\tr(L^{rot}(z))^2=I_{0,2}+I_{2,2}\wp(z)\,,~~
I_{0,2}=2\left(\frac{i}{2\pi \te}\right)^2 H\,,~~I_{2,2}=\tr\,\bfS^2\,.
$$

\begin{predl}\label{pr3.3}
The coefficients $I_{s,k}$ are in involution with respect to
the linear bracket (\ref{lb})
\beq{inv}
\{I_{s,k},I_{m,j}\}_1=0\,.
\eq
\end{predl}
{\sl Proof}\\
We have
$$
\{\tr(L_1^{rot}(z))^k,\tr(L_2^{rot}(w))^j\}_1=
\tr\{(L_1^{rot}(z))^k,(L_2^{rot}(w))^j\}_1\,.
$$
Then,
it follows from (\ref{LYB}) that these functionals  Poisson commute.
Using the expansion (\ref{8.21}) we arrive at the involutivity of the
coefficients (\ref{inv}).$\Box$
\bigskip

In particular, all functions $I_{s,k}$ Poisson commute with the
Hamiltonian $H$ (\ref{8.5}). Therefore,
they play the role of conservation laws of elliptic rotator
hierarchy on $\GLN$.
We have a tower of $\frac{N(N+1)}{2}$ independent integrals of motion
$$
\begin{array}{ccccc}
I_{0,2}&I_{2,2}&      &  & \\
I_{0,3}&I_{2,3}&I_{3,3}&  & \\
\ldots&\ldots&\ldots&\ldots&\\
I_{0,n}&I_{2,N}&\dots&\dots&I_{N,N}\\
\end{array}
$$
Note that $I_{k,k},~k=0,2,3\ldots,N$ are the Casimirs corresponding to the
coadjoint orbit
$$
{\mathcal R}^{rot}=\{\bfS\in\gln,~~\bfS=g^{-1}\bfS^{(0)}g\}\,.
$$
The conservation laws $I_{s,k}$ generate commuting flows on ${\cal R}^{rot}$
$$
\p_{s,k}\bfS=\{I_{s,k},\bfS\}_1\,, ~~(\p_{s,k}:=\p_{t_{s,k}})\,,
$$
that we call {\sl  ER$_N$ hierarchy}.

In what follows we will need another set of the conservation laws
coming from the
coefficients of the spectral curve
\beq{8.20}
{\cal C} ~:~ F(\la,z)\equiv\det(\la\cdot Id+L^{rot}(z))=0\,.
\eq
Here
$$
F(\la,z)=\la^N+\la^{N-2}a_2(z)+\ldots+a_N(z)\,,
$$
\beq{8.22}
a_k(z)=J_{0k}+J_{2k}\wp(z)+\ldots+J_{kk}\wp^{(k-2)}(z)\,.
\eq
Since $a_k(z)$ and $Tr(L(z))^k$ are related by the Newton formula,
one can recursively represent $J_{s,k}$ as polynomials of
 $I_{m,j},~j\leq k,~m\leq j$
\beq{JI}
J_{s,k}=\sum a_{\vec{l},\vec{m},\vec{j}}^{s,k}
\prod_{\sum j_im_i=k,\,l_i\leq j_i}(I_{l_i,j_i})^{m_i}\,,~~
(\vec{l}=(l_1,l_2,\ldots))\,.
\eq
For example,
$$
J_{s,2}=-I_{s,2}\,,~~(s=0,2)\,;~~~J_{s,3}=-I_{s,3}\,,~~(s=0,2,3)\,;
$$
\beq{JI1}
J_{0,4}=\frac{3}{2}I_{0,4}-I_{0,2}^2-\frac{g_2}{12}I_{2,2}^2\,,
~~(g_2=60\sum '_{m,n}(m+n\tau)^{-4})\,;
\eq
$$
J_{2,4}=\frac{3}{2}I_{2,4}-2I_{0,2}I_{2,2}\,;~~~J_{3,4}=\frac{3}{2}I_{3,4}
\;~~~J_{4,4}=\frac{3}{2}I_{4,4}-I^2_{2,2}\,,
$$
where in the second line we have exploited the relation $12\wp^2=2\wp''+g_2$.


\subsection{The classical $\GLN$ Feigin-Odesski algebras}

Here we consider the Feigin-Odesski Poisson brackets
related to $\GLN$. In notations of Ref.\,\cite{FO} it is the classical limit
of the quadratic associative algebra $Q_{N^2,k}$.

It turns out that the same $r$-matrix (\ref{6.1}) defines a quadratic
Poisson algebra.
We start with
the Lax operator $L^{rot}$ (\ref{8.7}). Modify it
in the following way
\beq{5.1}
L(z)=-S_0 Id+L^{rot}(z)\,,
\eq
where $S_0$ commute with $S_\al$ with respect to the linear brackets
(\ref{lb}).
Define the brackets between the entries of $L$ as follows
\beq{4.1}
\{L_1(z),L_2(w)\}_2=[r_{N,\te,\tau}(z-w),L_1(z)\otimes L_2(w)]\,,
\eq
(see Ref.\,\cite{Be,Skl}).
These brackets are
Poisson, since the
 Jacobi identity is provided by the
classical Yang-Baxter equation (\ref{YB}).
 A finite-dimensional quadratic Poisson algebra
is extracted from (\ref{4.1}) by
eliminating the dependence on the spectral parameters $z,w$.

\begin{predl}\label{pr3.4}
The quadratic Poisson algebra on $\gln^*$ has the form
\beq{1.1}
\{S_\al,S_0\}_2=
\sum_{\ga\neq\al}S_{\al-\ga}S_\ga(\wp_\te(\ga)-\wp_\te(\al-\ga))
\bfC_\te(\al,\ga)\,,
\eq
\beq{2.1}
\{S_\al,S_\be\}_2=S_0S_{\al+\be}
\bfC_\te(\al,\be)+\sum_{\ga\neq\al,-\be}S_{\al-\ga}S_{\be+\ga}
\bff_\te(\al,\be,\ga)\bfC_\te(\ga,\al-\be)\,,
\eq
where $\bfC_\te(\al,\be)$ are the $\sln$ structure constants (\ref{AA4}),
$$
\bff_\te(\al,\be,\ga)=
\ze_\te(\ga)+\ze_\te(\be-\al+\ga)
-\ze_\te(\be+\ga)+\ze_\te(\al-\ga)
$$
\beq{sc}
=-\frac{
\vth'(0)\vth_\te(\al)\vth_\te(\be)\vth_\te(\be-\al+2\ga)
}{
\vth_\te(\al-\ga)\vth_\te(\be+\ga)\vth_\te(\be-\al+\ga)\vth_\te(\ga)
}\,,
\eq
and $\vth_\te,~\ze_\te$ are the zeta constants (\ref{AA50}), (\ref{AA5}).
\end{predl}
{\sl Proof.}\\
The equation (\ref{4.1}) yields the following form of the brackets
\beq{4.1a}
\{S_a,S_b\}\vf_\te(a,z)\vf_\te(b,w)=
\sum_\ga S_{a-\ga}S_{b+\ga}\vf_\te(\ga,z-w)\vf_\te(a-\ga,z)\vf_\te(b+\ga,w)
\bfC_\te(\ga,a-b)\,.
\eq
Note first, that the common exponents in the expressions of the functions
$\vf_\te$ (\ref{vf}) coincide in the left and the right sides.
This allows one to pass to the functions
$\phi_\te$ and in this way to use the Fay-type identities.
We rewrite the last expression in the form
$$
\{S_a,S_b\}\phi_\te(z)\phi_\te(w)=
\sum_\ga S_{a-\ga}S_{b+\ga}\bfC_\te(\ga,a-b)
$$
$$
\times\left(
\phi_\te(\ga,z-w)\phi_\te(a-\ga,z)\phi_\te(b+\ga,w)
-\phi_\te(a-b-\ga,z-w)\phi_\te(b+\ga,z)\phi_\te(a-\ga,w)\right)
\,.
$$
Consider first the case $b=0$.
Then we can apply (\ref{ir2}) for $u=(a_1+a_2\tau)\te$, $v=(\ga_1+\ga_2)\te$.
It immediately leads us  to (\ref{1.1}).
If $a,b\neq 0$ we have the terms of two types.
For $\ga\neq a,-b$ one uses (\ref{ir}), (\ref{ir1}) and finds
the second term in the right
side of (\ref{2.1}). If $\ga=a$ or $\ga-=b$ one should take into
account that $\phi_\te(0,z)=1$ and then use
the Fay identity (\ref{ad3}).  This  gives us the first term.
The identity (\ref{sc}) follows from (\ref{ir3}). $\Box$

\bigskip

We denote by $P_{N,\te,\tau}$ the quadratic Poisson
algebra (\ref{1.1}), (\ref{2.1}). Recall
that $\tau$ is the modular parameter of the auxiliary curve $E_\tau$
and $\te=k/N$. The algebra $P_{2,\oh,\tau}$ is
 the classical Sklyanin algebra \cite{Skl}. In this case the r.h.s. of
(\ref{2.1}) contains only the first term.

As for the linear brackets, there exist
$N$ Casimirs $C_0^{(2)},C_2^{(2)},\ldots,C_N^{(2)}$
of the algebra $P_{N,\te,\tau}$
\cite{BDOZ}.
They can be read off from the expansion of the spectral curve (\ref{8.20})
in the basis of elliptic functions
\beq{7.1}
\det(S_0\cdot Id+L^{rot}(z))=C_0^{(2)}+\sum_{k=2}^NC_k^{(2)}\wp^{(k-2)}(z)
\,.
\eq
It follows from (\ref{8.20}), (\ref{8.22}) and (\ref{5.1}) that they are
linear combinations of the integrals $J_{s,k}$
\beq{8.1}
C_0^{(2)}=S_0^N+\sum_{m=2}^NS_0^{N-m}J_{0,m}\,,
\eq
\beq{8a.1}
C_k^{(2)}=\sum_{m=0}^{N-k}S_0^{N-k-m}J_{k,k+m}\,,~~(k=2,\ldots,N)\,.
\eq


\subsection{The bihamiltonian structure}

Two Poisson structures are called {\sl compatible} (or, form {\sl
a Poisson pair}) if their linear combinations are Poisson structures as well.

\begin{predl}\label{pr3.5}
The linear and quadratic Poisson brackets on $\gln$ are compatible.
\end{predl}
{\sl Proof}\\
Let us replace $S_0\to S_0+\la$ in  the quadratic brackets (\ref{1.1}),
(\ref{2.1}).
Then (\ref{1.1}) does not change, while  (\ref{2.1}) acquires an
additional linear
term. In this way we define a  one-parameter family of the brackets
\beq{9.1}
\{\bfS,\bfS\}_\la:=\{\bfS,\bfS\}_2+\la\{\bfS,\bfS\}_1\,,
\eq
where $\{\bfS,\bfS\}_1$ denote the linear brackets (\ref{lb}).
Therefore, the linear combination
$\{\bfS,\bfS\}_\la$ of two brackets can be obtained from the quadratic
bracket by a simple shift of $S_0$. The result of the shift,
of course, still satisfies
the Jacobi identity, and hence it is a Poisson bracket for any $\la$.
$\Box$
\bigskip

We denote this family of the quadratic Poisson algebras by
$P_{N,\te,\tau,\la}$.
The algebras are isomorphic for different $\la$ and degenerate to
the linear Poisson algebra $\gln^*$ as $\la\to\infty$.

Consider the Casimir functions
 of the Poisson algebra $P_{N,\te,\tau,\la}$
$$
h_k(\la)=C_k^{(2)}(S_0+\la)\,~~(k=0,2,\ldots,N)\,.
$$
It follows from (\ref{8.1}) and (\ref{8a.1}) that $h_k(\la)$ are
polynomials in $\la$
\beq{11.1}
h_k(\la)=\sum_{m=0}^{N-k}h_{m,k}\la^m\,.
\eq
The coefficients are defined as
$$
h_{a,k}=\f1{a!}\p_\la^{(a)}C_k^{(2)}(S_0+\la)_{\la=0}\,.
$$
It implies that
\beq{12a.1}
h_{a,0}=\frac{N!}{(N-a)!a!}S_0^{N-a}+
\sum_{m=2}^{N-a}\frac{(N-m)!}{(N-m-a)!a!}S_0^{N-m-a}J_{0,m}\,,
\eq
\beq{12b.1}
h_{a,k}=
\sum_{m=0}^{N-k-a}\frac{(N-k-m)!}{(N-k-m-a)!a!}S_0^{N-k-m-a}J_{k,k+m}\,,
(k=2,\ldots,N)\,.
\eq
In particular,
\beq{13.1}
h_{N-1,0}=NS_0\,,~~h_{N-2,0}=\frac{N(N-1)}{2}S_0^2+J_{0,2}\,.
\eq
 Conversely, one can express $J_{m,k+m}$ as a linear
combination of $h_{N-m-s,m},~s=0,\ldots,k$ from
the relations (\ref{12a.1}) and (\ref{12b.1}). For example,
$$
J_{j,j}=h_{N-j,j}\,,~~J_{j,j+1}=h_{N-j-1,j}-(N-j)S_0h_{N-j,j}\,,
$$
$$
J_{j,j+2}=h_{N-j-2,j}-(N-j)S_0h_{n-j-1,j}
+\frac{(N-j)(N-j-1)}{2}S_0^2h_{N-j,j}\,.
$$
We arrange the quantities $h_{m,k}$ in the triangular tableau
\bigskip
\beq{ta}
\begin{array}{ccccccccccccc}
h_{0,0} & h_{1,0}& &\ldots  &\ldots&\ldots&\ldots&\ldots&\ldots&\ldots&&h_{N-1,0} & h_{N,0}\\
        & h_{0,2} & &h_{1,2}& &\ldots&\ldots&\ldots&  &h_{N-3,2}&  &h_{N-2,2}&\\
        &     &h_{0,3}&   &h_{1,3}&\ldots&\ldots&\ldots        &h_{N-4,4}&  &h_{N-3,3}&      &    \\
   &&&  \ldots&        &\ldots&\ldots                    &\ldots&  &\ldots&&&\\
   &&&    & h_{0,N-1}&  & &   &h_{1,N-1}& &&&    \\
     &&&&   &          & h_{0,N}&   &&&&&
\end{array}
\eq
\bigskip

Note, that the second line corresponding to $h_{s,1}$ is absent
since $h_1(\la)=0$.
The left side of the triangle  contains the Casimirs
$h_{0,j}=C^{(2)}_j$ (\ref{8.1}), (\ref{8a.1}) of $P_{N,\te,\tau}$,
while the right side represents
the Casimirs $h_{N-j,j}=J_{j,j}$ of the linear brackets on $\gln^*$.

The remarkable property of the quantities $h_{a,k}$ is that they are
in involution with respect to both the linear and quadratic  brackets:
$$
\{h_{a,k},h_{b,j}\}_{1,2}=0\,,
$$
see, e.g., \cite{Mag}. This fact follows from the identities
\beq{10.1}
\{h_k(\la),h_j(\la)\}_\la=0\,,
\eq
held for each $\la$.

\begin{predl}\label{pr3.6}
The integrals $h_{a,k}$ provide the recurrence relation
 for the elliptic rotator hierarchy $ER_N$
\beq{12.1}
\{h_{a,k},\bfS\}_1=-\{h_{a+1,k},\bfS\}_2\,.
\eq
\end{predl}

{\sl Proof.}\\
Since $h_k(\la)$ are the Casimirs, we have
$\{h_k(\la),\bfS\}_\la=0$.
Substitute the representation (\ref{11.1}) into this equation.
Then the recurrence relation (\ref{12.1})
comes from the coefficient in front of $\la^a$.
$\Box$
\bigskip

In this way one can start with the Casimirs of the linear brackets
and produce a non-trivial dynamical system using the quadratic brackets
\beq{13a.1}
\{J_{k,k},\bfS\}_2=-\{J_{k,k+1},\bfS\}_1\,,
\eq
(see (\ref{13.1})). In particular,
 the flow (\ref{8.5a}) corresponding to $H=I_{0,2}$
can be represented by the quadratic brackets with $S_0$:
$$
\p_t\bfS=\{S_0,\bfS\}_2\,.
$$
Alternatively, one can start with the Hamiltonians in the left side
of the table (\ref{ta}) and the linear brackets.

\bigskip
This allows us to conclude the proof of Theorem 2.1 for the hierarchy ER$_N$.
Namely, Propositions \ref{pr3.2}, \ref{pr3.4}, \ref{pr3.5}, and \ref{pr3.6} are
equivalent, respectively, to {\bf i)}, {\bf ii)}, {\bf iii)}, and {\bf iv)}.


\section{Elliptic Rotators on non-commutative torus}
\setcounter{equation}{0}

\subsection{The Lax equation and integrals of motion}

{\sl 1. The system description}\\
The elliptic rotator on the non-commutative torus
$\cA_\te$ is a  generalization
of the elliptic rotator on $\GLN$.
We consider the Euler-Arnold top  on the group
$ SIN_\te$ of the NCT algebra $\cA_\te$ (Appendix C).
The Lie coalgebra $sin^*_\te$ is equipped
with the linear Poisson brackets
\beq{pb1}
\{S_\al,S_\be\}_1=\bfC_\te(\al,\be)S_{\al+\be}\,,~~(\al\in\ti{\mZ}^{(2)})\,,
\eq
where $\bfS=\sum S_{-\al} T_\al$ and
$$
\ti{\mZ}^{(2)}=\{\al=(\al_1,\al_2)\,,~\al_j\in\mZ\,,~\al\neq (0,0)\}\,.
$$

The inverse inertia tensor $\bfJ$ maps
$sin_\te^*\to\sin_\te$, and
depends on four parameters: $\te,\tau\in\mC,\,(\Im m\tau>0)$ and $\ep=(\ep_1,\ep_2)$,
where $0<\te\ep_a\leq 1$ and $\ep_a\te$ are irrational numbers. The
components of $\bfJ$ are given by
the elliptic constants $J_\al=\wp_{\te,\ep}(\al)$
(\ref{C1})
$$
\bfJ(\bfS)=\sum_{\al\in\ti{\mZ}^{(2)}} \wp_{\te,\ep}(\al) S_{-\al} T_\al\,.
$$
The Hamiltonian has the form
\beq{4.3}
H=2\pi^2\te^2\int_{\cA_\te}\bfS \cdot\bfJ(\bfS)=
-\frac{1}{2}\sum_{\al\in\ti{\mZ}^{(2)}} \wp_{\te,\ep}(\al) S_{\al} S_{-\al}\,.
\eq

Introduce a complex structure on the NCT $\cA_\te$ depending on
$\ep=(\ep_1,\ep_2)$ and $\tau$  such that for $X=\sum_a c_aT_a$
one has:
\beq{cs}
\bp_{\ep,\tau} X=\sum_a (\ep_1a_1+\ep_2 a_2\tau)c_aT_a\,.
\eq
Then, in the Moyal representation, the operator  $\bfJ$ can be identified
with the pseudo-differential operator
\beq{J}
\bfJ(\bfS)(x)=\wp(\te\bp_{\ep,\tau})\bfS(x)\,,
\eq
and
\beq{4.3a}
H=2\pi^2\te^2\int_{\cA_\te}\bfS \cdot \wp(\te\bp_{\ep,\tau})\bfS\,.
\eq
The equation of motion in the form of the Moyal brackets has the
standard form
\beq{4.4}
\p_t\bfS=\{\bfS,(\wp(\te\bp_{\ep,\tau})\bfS)\}^\te:=\ad^*_{\bfJ(\bfS)}\bfS\,,
\eq
or, in the components,
\beq{4.2}
\p_tS_\al =\sum _{\ga\in\ti{\mZ}^{(2)}} S_\ga\wp_{\te,\ep}(\ga) S_{\al-\ga}\,.
\eq

\bigskip
\noindent
{\sl 2. The $r$-matrix and the Lax equation}\\
Define formally the Lax operator
\beq{4.9}
L^{rot}(z)=-\sum_{\al\in\ti{\mZ}^{(2)}} S_{\al} \vf_{\te,\ep}(\al,z)T_\al\,,
\eq
where $\vf_{\te,\ep}$ is given by (\ref{CC2}).
Below we formulate the conditions
on the phase space, which show that $L^{rot}(z)$ is well defined
for $z\in E_\tau,~z\neq 0$. Note that
$$
\int_{\cA_\te}
L^{rot}(z)=0\,.
$$

The following Proposition is an infinite-dimensional analog of
Proposition \ref{pr3.1}.
\begin{predl}\label{pr4.1}
The equations of motion (\ref{4.2}) have the Lax form
\beq{4.5}
\p_tL^{rot}=[L^{rot},M^{rot}]:=\{L^{rot},M^{rot}\}^\te
\eq
with
\beq{4.9d}
M^{rot}(z)=-\sum_{\al\in\ti{\mZ}^{(2)}} S_{\al} f_{\te,\ep}(\al,z)T_\al\,,
\eq
 and $f_{\te,\ep}$
(\ref{CC3a}).
\end{predl}
The proof is analogous to the finite-dimensional case.
It is based on the Calogero functional equation
(\ref{ad2}).
\bigskip

Introduce the classical $r$-matrix on $sin_\te\otimes sin_\te$
as the following sum:
\beq{1.20}
r_{\te,\ep,\tau}(z-w)
=\sum_{\ga\in\ti{\mZ}^{(2)}}\vf_{\te,\ep}(\ga,z-w)T_\ga\otimes T_{-\ga}\,,
\eq
Lemma \ref{le3.1} and Proposition \ref{pr3.2} have the following analogs
for the NCT:
\begin{predl}\label{pr4.2}
$a)$ The $r$-matrix (\ref{1.20}) satisfies the classical
Yang-Baxter equation
(\ref{YB}).\\
$b)$ The $r$-matrix defines the Poisson brackets
for the entries of the  Lax operator (\ref{4.9})
\beq{LYB1}
\{L_1^{rot}(z),L_2^{rot}(w)\}_1=[r_{\te,\ep,\tau}(z-w),L^{rot}(z)\otimes Id+
Id\otimes L^{rot}(w)]\,.
\eq
They are equivalent to the linear brackets (\ref{pb1}) on $\sin^*_\te$.
\end{predl}
The proof of Proposition \ref{pr4.2}$(a,b)$ is similar to those of
Lemma \ref{le3.1} and Proposition \ref{pr3.2} and  is based on the
Fay identity (\ref{ad3}).

\bigskip
\noindent
{\sl 3. The hierarchy of the Lax equations}\\
In order to construct integrals of motion we first discuss the properties
of the Lax operator.
This  operator is the meromorphic quasi-periodic function
on $E_\tau$ with a simple pole
$$
Res\, L^{rot}(z)|_{z=0}=\bfS\,,
$$
$$
L^{rot}(z+1)=U_1^{\ep_1}L^{rot}(z)U_1^{-\ep_1}\,,~~
L^{rot}(z+\tau)=U_2^{\ep_2} L^{rot}(z)U_2^{-\ep_2}\,.
$$
where $U_1,U_2$ are the generators of NCT $\cA_\te$ (\ref{3.1}).
(This follows from (\ref{4.9}), (\ref{A.3a}) and (\ref{qp}).)
It means that the integrals
$$
\int_{\cA_\te}(L^{rot}(z))^j
$$
are doubly periodic functions on $E_\tau$ with poles up
to order $j$.  Thereby, we have the expansion
\beq{4.12}
\int_{\cA_\te} (-L^{rot}(z))^j=I_{0,j}+
\sum_{r=2}^{j}I_{r,j}\wp^{(r-2)}(z)\,.
\eq
In particular,
$$
\int_{\cA_\te} (L_\te)^2(z)=I_{0,2}+
\wp(z)\int_{\cA_\te}\bfS^2\,,~{\rm where }
~I_{0,2}=2\left(\frac{i}{2\pi \te}\right)^2H\,.
$$
Note that
$$
I_{j,j}=\int_{\cA_\te}\bfS^j
$$
 are the Casimirs of the linear brackets.
\bigskip

{\sl The phase space} $\cR$ of the elliptic rotator is determined by the
following properties of the integrals $I_{s,j}$
\beq{ps}
\cR=\{\bfS\in \cA_\te^*~|~i)~I_{s,j}<\infty\,,~~
ii)\lim_{j\to\infty}\frac{I_{s,j+1}(\bfS)}{I_{s,j}(\bfS)}
< 1~{\rm for ~all~} s\leq j\}\,.
\eq
The first condition means that the traces (\ref{4.12}) of the
Lax operator are well defined on $E_\tau$ for $z\neq 0$.
 In particular, we have
\beq{ps1}
I_{0,2}<\infty\,,~{\rm i.e.} ~\sum \wp_\te(\al) S_{\al} S_{-\al}<\infty\,.
\eq
We will use the second condition below
to define the Casimir functions for the family of quadratic Poisson brackets.
\bigskip

Further, due to Lemma \ref{le3.1},  an analog of Proposition \ref{pr3.2}
is the following.
\begin{predl}
The quantities $I_{s,k}$ are pairwise Poisson commute.
\end{predl}
The conservation laws $I_{s,k}$ generate commuting flows on the
phase space $\cR$ with respect to the linear brackets
\beq{erh}
\p_{s,k}\bfS=\{I_{s,k},\bfS\}^\te_1\,,
\eq
 where $\p_{s,k}$ stands for the corresponding time
derivative $\p_{t_{s,k}}$.
We call these equations {\sl $ER_\theta$ hierarchy on the NCT}.

 One can show that all flows can be represented in the Lax form
with $L^{rot}$ (\ref{4.9}) and the corresponding $M_{s,k}$.
Furthermore, there exists a set of the integrals of $J_{s,k}$-type.
They are related to the integrals
$I_{m,j}$ by the same formulae (\ref{JI}), (\ref{JI1}).


\subsection{Quadratic Poisson algebras and the bihamiltonian structure
on NCT}

To define the quadratic Poisson algebra on the phase space
$\cR$ (\ref{ps}) we modify the Lax operator
(\ref{4.9}):
\beq{4.9a}
L=-S_0\cdot T_0+L^{rot}=S_0\cdot T_0+\sum_\al
S_{-\al}\vf_{\te,\ep}(\al,z)T_\al\,,
\eq
where $S_0$ commutes with $S_\al$ with respect to the linear brackets.
Due to the YB equation, the
$r$-matrix (\ref{1.20}) defines the quadratic Poisson brackets
\beq{4.9b}
\{L_1(z),L_2(w)\}^\te_2=[r_{\te,\ep,\tau}(z-w),L_1(z)\otimes L_2(w)]\,.
\eq
Again, we can get rid of  the dependence on the spectral parameter.
\begin{predl}
The quadratic Poisson brackets on the phase space $\cR$ are defined
as follows
\beq{1.10}
\{S_\al,S_0\}^\te_2=
\sum_{\ga\neq\al}S_{\al-\ga}S_\ga(\wp_{\te,\ep}(\ga)-\wp_{\te,\ep}(\al-\ga))
\bf\bfC_\te(\al,\ga)\,,
\eq
\beq{2.10}
\{S_\al,S_\be\}_2^\te=S_0S_{\al+\be}
\bfC_\te(\al,\be)+\sum_{\ga\neq\al,-\be}S_{\al-\ga}S_{\be+\ga}
\bff_{\te,\ep}(\al,\be,\ga)\bfC_\te(\ga,\al-\be)\,,
\eq
where $\bfC_\te(\al,\ga)$ is a structure constant of the $sin_\te$-algebra,
\beq{2a.10}
\bff_{\te,\ep}(\al,\be,\ga)=
\ze_{\te,\ep}(\ga)+\ze_{\te,\ep}(\be-\al+\ga)
-\ze_{\te,\ep}(\be+\ga)+\ze_{\te,\ep}(\al-\ga)
\eq
$$
=-\frac{
\vth'(0)\vth_{\te,\ep}(\al)\vth_{\te,\ep}(\be)\vth_{\te,\ep}(\be-\al+2\ga)
}{
\vth_{\te,\ep}(\al-\ga)\vth_{\te,\ep}(\be+\ga)\vth_{\te,\ep}(\be-\al+\ga)\vth_{\te,\ep}(\ga)
}\,,
$$
and $\vth_{\te,\ep},~\ze_{\te,\ep}$ are the constants (\ref{C0}), (\ref{C1}).
\end{predl}
{\sl Proof}\\
  The equality (\ref{4.9b}), being reduced to the coefficients
in the front of $T_a\otimes T_b$,
can be rewritten in the form (\ref{4.1a}) where
the functions $\vf_{\te}(a,z)$ are replaced on $\vf_{\te,\ep}(a,z)$.
Then using the cubic functional equations (\ref{ir}) and (\ref{ir1}) we obtain
to the algebra (\ref{1.10}), (\ref{2.10}).
 Due the definition of the phase space (\ref{ps1})
the series in the right hand sides of (\ref{1.10}), (\ref{2.10}) converge.
$\Box$

\bigskip
We denote this Poisson algebra by $\cP_{\te,\ep,\tau}$. Set now
$\ep_1=\ep_2=1$.
The corresponding algebra $\cP_{\te,1,\tau}$ can be considered as a special
large $N$ limit of the finite-dimensional algebras
$P_{N,\te,\tau}$ (\ref{1.1}), (\ref{2.1}). For this we just replace
the rational number $\te=k/N$ by an arbitrary irrational
number $0\leq\te<1$ in the algebra $P_{N,\te,\tau}$.

\bigskip
\begin{predl}
$a)$ The linear (\ref{pb1}) and quadratic (\ref{1.10}), (\ref{2.10})
 Poisson brackets defined on ${\cal R}$ are compatible.\\
$b)$ The  $ER_\te$ hierarchy admits the bihamiltonian structure.
\end{predl}
{\sl Proof.}\\
One can shift $S_0+\la$ and define the family Poisson algebras
$P_{\te,\ep,\tau,\la}$.
To define the Casimir elements on $P_{N,\te,\tau,\la}$ we consider
$$
\log\det(S_0+\la+L^{rot}(z))=
\log(S_0+\la)+\sum_{k=2}^\infty\frac{(-1)^{k+1}}{k(S_0+\la)^k}\tr L(z)^k\,.
$$
Then we come to the infinite set of  Casimirs
\footnote{Since we pass from $\det$ to $\log\det$ the basis of the Casimir
functions differs
from (\ref{8.1})}
$$
\ti{C}_0^{(2)}(S_0+\la)=
\log(S_0+\la)+\sum_{k=2}^\infty\frac{(-1)^{k+1}}{k(S_0+\la)^k}I_{0,k}\,,
$$
$$
\ti{C}_j^{(2)}(S_0+\la)
=\sum_{k=2}^\infty\frac{(-1)^{k+1}}{k(S_0+\la)^k}I_{j,k}\,,~
(j=2,3,\ldots)\,.
$$
These functionals are well-defined on the phase space (\ref{ps}).
This allows us to introduce the new set of conserved quantities
$$
\ti{C}_0^{(2)}(S_0+\la)=\sum_s\ti{h}_{s,0}\la^s\,,
$$
$$
\ti{C}_j^{(2)}(S_0+\la)=\sum_s\ti{h}_{s,j}\la^s\,.
$$
The latter Poisson commute with respect to the both types of
brackets and give rise to the bihamiltonian structure of the
hierarchy
\beq{bh}
\{\ti{h}_{s,j},\bfS\}^\te_1=-\{\ti{h}_{s+1,j},\bfS\}^\te_2\,.
\eq
In particular,  we can represent the flow (\ref{4.4}) in the form
$$
\p_t\bfS=\{S_0,\bfS\}_2^\te\,.
$$
(see (\ref{13.1})). $\Box$

This proposition concludes the proof of Theorem \ref{main-th} for the
hierarchy $ER_\te$.


\section{Elliptic Rotators on $SDiff(T^2)$}\label{DIFF-ER}
\setcounter{equation}{0}

\subsection{Description of the hierarchy}

In the "dispersionless" limit $\te\to 0$ the Lie algebra $sin_\te$
turns to the Lie algebra $Ham(T^2)$ of Hamiltonians on a two-dimensional
torus, see (\ref{Ham}). This  algebra of Hamiltonian functions
can be represented by the Lie algebra of the
corresponding divergence-free vector fields $SVect(T^2)$.
More precisely, to pass from $Ham(T^2)$ to $SVect(T^2)$
one has to discard the constant Hamiltonians, but add the ``flux''
vector fields $\partial/\partial x_1$ and $\partial/\partial x_2$
corresponding to multivalued Hamiltonian functions
$x_1$ and $ x_2$ on the torus.

We define the elliptic rotator system $ER$
on the Lie group $SDiff(T^2)$ by
considering the limit $\te\to 0$ of the $ER_\te$-system described above.
Let
$\te\to 0,~\ep_{1,2}\to\infty$,
such that
\beq{rs}
\lim_{\te\to 0}(\te\ep_{1,2})=\ep'_{1,2}<1\,,~
~\ep'_{1,2}~{\rm are~irrational}\,.
\eq
In what follows we drop the superscript $'$.

Let $\bfS=\sum_\al S_{-\al}\bfe(\al\cdot x)\in Ham^*(T^2)$,
where $\bfe(\al\cdot x)$ is the Fourier basis (\ref{fb}) of $Ham^*(T^2)$.
In  the Fourier basis, the linear Poisson bracket assumes the form
\beq{3.7a}
\{S_\al,S_\be\}_1=(\al\times\be)S_{\al+\be}\,,~\al\times\be=\al_1\be_2-\al_2\be_1\,.
\eq

The inverse
inertia operator $\bfJ\,:\,Ham^*(T^2)\to Ham(T^2)$  becomes
$$
\bfJ\,:~S_\al \to \wp_{\ep}(\al) S_\al\,,~~
\wp_{\ep}(\al)=\wp(\ep_1\al_1+\ep_2\al_2\tau;\tau)\,,
$$
where $\al\in\ti{\mZ}^{(2)}$ (\ref{lat}).
The operator is well defined since $\ep_j$ are irrational.
In other words, the operator $\bfJ$ is the pseudo-differential operator
$$
\bfJ\,:~\bfS(x) \to \wp(\bp_{\ep,\tau})\bfS(x)\,,
$$
where $\bp_{\ep,\tau}$ is the operator of the complex structure on the
elliptic curve, commutative torus $T^2$.
In fact, the complex structure depends on the ratio
$\tau{\ep_2}/{\ep_1}$.

The quadratic Hamiltonian of the system is
\beq{1.7}
H=-\oh\sum_\ga S_\ga\wp_{\ep}(\ga)S_{-\ga}=-\oh
\int_{T^2}\bfS(\wp(\bp_{\ep,\tau})\bfS)\,.
\eq
and the corresponding equations of motion are
\beq{2.7}
\p_t\bfS=\{\bfS,\wp(\bp_{\ep,\tau})\bfS\}_1\,.
\eq
This equation is  an elliptic analog of the hydrodynamics
equation on the torus $T^2$, regarded as an elliptic curve, see Remark below.
We call it {\sl the elliptic hydrodynamics}.

Define the Lax operator
\beq{3.7}
L^{rot}(x;z)=-\sum_{\al\in\ti{\mZ}^{(2)}} S_{\al}\vf_{\ep}(\al,z)\bfe(\al\cdot
x)\,,~~
\vf_{\ep}(\al,z)=\vf(\ep_1\al_1+\tau\ep_2\al_2,z)\,.
\eq
The conditions on the phase space formulated below, see (\ref{ps0}),
ensure that the
operator $L^{rot}(x,z)$ is well defined
for $z\in E_\tau,~z\neq 0$. Note that
$$
\int_{T^2}
L^{rot}_{\ep}(z)=0\,.
$$

\begin{predl}\label{pr5.1}
The equations of motion (\ref{2.7})
 have the dispersionless  Lax
representation
$$
L^{rot}=\{L^{rot},M^{rot}\}_1
$$
with $L^{rot}$ given by (\ref{3.7}) and
\beq{4.7}
M^{rot}(x;z)=-\sum_{\al\in\ti{\mZ}^{(2)}} S_{\al}f_{\ep}(\al,z)\bfe(\al\cdot
x)\,,
\eq
where
$$
f_{\ep}(\al,z)=f(\ep_1\al_1+\tau\ep_2\al_2,z)=
\bfe(\ep_2\al_2)\p_u\phi(u,z)|_{u=\ep_1\al_1+\tau\ep_2\al_2}\,.
$$
\end{predl}
The proof of Proposition \ref{pr5.1} is the same as Proposition \ref{pr3.1}.

Note that operators $L^{rot}$ and $M^{rot}$ satisfy the quasi-periodicity
properties
\beq{5.7}
L^{rot}(x_1,x_2;z)=L^{rot}(x_1+\ep_2,x_2;z+1)\,,
\eq
\beq{6.7}
L^{rot}(x_1,x_2;z)=L^{rot}(x_1,x_2-\ep_1;z+\tau)\,,
\eq
\beq{7.7}
M^{rot}(x_1,x_2;z)=M^{rot}(x_1+\ep_2,x_2;z+1)\,,
\eq
and
\beq{8.7b}
M^{rot}(x_1,x_2;z)-M^{rot}(x_1,x_2-\ep_1;z+\tau)=
2\pi i L^{rot}(x_1,x_2,z)\,.
\eq
Furthermore, it follows from (\ref{5.7}) and (\ref{6.7}) that there exists
the expansion
\beq{12.7}
\int_{T^2}
(L^{rot}(x,z))^k=I_{0,k}+\sum_{s=2}^kI_{s,k}\wp^{(s-2)}(z)\,.
\eq

{\sl The phase space} $\cR^0$ of the elliptic rotator on $SDiff(T^2)$
can be described similarly to the phase space $\cR$ for $SIN_\te$-group, see
(\ref{ps}):
\beq{ps0}
\cR^0=\{\bfS\in Ham^*(T^2)\,| ~i)~I_{s,j}<\infty\,,~~
ii)\lim_{j\to\infty}\frac{I_{s,j+1}(\bfS)}
{I_{s,j}(\bfS)}<1~{\rm for~all}~s\leq j\}\,.
\eq

Define the $r$-matrix on $T^2\otimes T^2$ by
\beq{10.7}
r_{\ep,\tau}(x,y;z)=
\sum_{\al\in\ti{\mZ}^{(2)}}\vf_{\ep}(\al,z)\bfe(\al\cdot x)\bfe(-\al\cdot y)\,,~z\in E_\tau\,.
\eq
\begin{predl}\label{le5.1}
$a)$ The $r$-matrix (\ref{10.7}) satisfies the "dispersionless"
YB equation
$$
\{r_{\ep,\tau}(x,y;z-w),r_{\ep,\tau}(x,v;z)\}
+\{r_{\ep,\tau}(x,y;z-w),r_{\ep,\tau}(y,v;w)\}
$$
\beq{DYB}
+
\{r_{\ep,\tau}(x,v;z),r_{\ep,\tau}(y,v;w)\}=0\,.
\eq
$b)$ In terms of the Lax operator (\ref{3.7})
the canonical brackets on $T^2$ take the form
\beq{11.7}
\{L^{rot}(x;z),L^{rot}(y;w)\}=
\{r_{\ep,\tau}(x,y,z-w),L^{rot}(x;z)\}+\{r(x,y,z-w)
L^{rot}(y;w)\}\,.
\eq
\end{predl}
{\sl Proof}\\
These statements are consequences of the Fay identity
(\ref{ad3}), cf. Proposition \ref{pr4.2}.$\Box$
\bigskip

The form of the brackets (\ref{11.7}) implies that
$$
\{\int_{T^2}
(L^{rot}(x,z))^k,\int_{T^2}
(L^{rot}(x,z))^j\}=0\,.
$$
In turn, then (\ref{12.7}) produces  the infinite sequence
of conservation laws $I_{s,k}$ in involution.
They define the hierarchy of commuting flows
\beq{13.7}
\p_{s,k}\bfS=\{I_{s,k},\bfS\}_1\,,
\eq
which is the dispersionless limit of the hierarchy
(\ref{erh}). The equations can be represented in the form of the
dispersionless Lax equations.


\subsection{Quadratic Poisson algebras, and the bihamiltonian structure
on $SDiff(T^2)$}

To describe the quadratic Poisson brackets, we first
pass from the Lie algebraic Lax operator $L^{rot}$ to the operator
\beq{14.7}
L_{\ep}=-S_0+L^{rot}(x;z)\,.
\eq

Define the quadratic Poisson algebra by the formula
\beq{15.7}
\{L_{\ep}(x;z),L_{\ep}(y;w)\}=
\{r_{\ep,\tau}(x,y;z-w),L_{\ep}(x;z) L_{\ep}(y;w)\}\,.
\eq

\begin{predl}
In terms of the Fourier modes, the Poisson  algebra (\ref{15.7}) has the form
\beq{16.7}
\{S_\al,S_0\}_2=
\sum_{\ga\neq\al}S_{\al-\ga}S_\ga(\wp_{\ep}(\ga)-\wp_{\ep}(\al-\ga))
(\al\times\ga)\,,
\eq
\beq{17.7}
\{S_\al,S_\be\}_2=S_0S_{\al+\be}
(\al\times\be)+\sum_{\ga\neq\al,-\be}S_{\al-\ga}S_{\be+\ga}
\bff_{\ep}(\al,\be,\ga)(\ga\times(\al-\be))\,,
\eq
where
$$
\bff_{\ep}(\al,\be,\ga)=
\ze_{\ep}(\ga)+\ze_{\ep}(\be-\al+\ga)
-\ze_{\ep}(\be+\ga)+\ze_{\ep}(\al-\ga)
$$
$$
=-\frac{
\vth'(0)\vth_{\ep}(\al)\vth_{\ep}(\be)\vth_{\ep}(\be-\al+2\ga)
}{
\vth_{\ep}(\al-\ga)\vth_{\ep}(\be+\ga)\vth_{\ep}(\be-\al+\ga)\vth_{\ep}(\ga)
}\,,
$$
$\ze_{\ep}(\ga)=\ze(\ep_1\ga_1+\tau\ep_2\ga_2)$, $\vth_{\ep}(\ga)=\ze(\ep_1\ga_1+\tau\ep_2\ga_2)$.
\end{predl}
The proof of this Proposition is similar to that of Proposition \ref{pr3.2}.

By the same trick as before we can show that the linear and the quadratic
brackets are
compatible. It allows us to construct the integral of the $\ti{h}_{s,k}$ type
and define the recurrence representation (\ref{bh}) for
the bihamiltonian structure on the hierarchy. In particular, the
equation (\ref{2.7}) can be represented    as
$$
\p_t\bfS=\{S_0,\bfS\}_2\,.
$$

\subsection{Rational limit of the elliptic hydrodynamics}

{\sl 1. Description of the limit}\\
So far we have been dealing with the elliptic curve $E_\tau$
parameterized by the two half-periods $\om_1,\om_2,\,\tau=\om_2/\om_1$.
WE replace our main functions in the following way:
\beq{5.10}
\ze_\ep(\ga)=\ze(\ep_1\ga_1\om_1+\ep_2\ga_2\om_2;\om_1,\om_2,)\,,~~
\wp_\ep(\ga)=\wp(\ep_1\ga_1\om_1+\ep_2\ga_2\om_2;\om_1,\om_2,)\,.
\eq
Now we are going to consider the
rational limit of the elliptic curves, i.e., degeneration
$\lim\om_{1,2}\to\infty$. We look at the double scaling limit
$$
\lim\ep_{1,2}\to 0\,,~\lim\om_{1,2}\to\infty\,,~
\lim\ep_1\om_1=1\,,~\lim\ep_2\om_2=\tau\,.
$$
Then $\lim\om_{1,2}\to\infty$ leads to the rational degeneration
of the elliptic functions
(\ref{5.10})
$$
\lim_{\om_{1,2}\to\infty}\ze_\ep(\ga)=\f1{\ga_\tau}\,,~~
\lim_{\om_{1,2}\to\infty}\wp_\ep(\ga)=\f1{\ga_\tau^2}\,,~~
\ga_\tau=\ga_1+\ga_2\tau
$$
This implies, in particular, that the inverse inertia tensor assumes the form
\beq{it}
\bfJ(\bfS)=\bp^{-2}\bfS(x_1,x_2)
=\sum_{\al\in\ti{\mZ}^{(2)}} \f1{\al_\tau^2} S_{-\al} T_\al\,,
\eq
where $\al_\tau=\al_1+\al_2\tau$.
The equations of motion are defined by the corresponding quadratic Hamiltonian
\beq{qh}
H=-\frac{1}{2}\int_{T^2}\bfS \bp^{-2}\bfS=
-\frac{1}{2}\sum_{\al\in\ti{\mZ}^{(2)}} \f1{\al_\tau^2} S_{\al} S_{-\al}\,.
\eq
with respect to the linear Poisson bracket on $Ham(T^2)$:
\beq{cem}
\p_t\bfS=\{\bfS,\bp^{-2}\bfS)\}_1\,.
\eq
In the Fourier components the latter becomes
$$
\p_tS_\al =\sum _{\ga\in\ti{\mZ}^{(2)}} S_\ga \f1{\ga_\tau^2}S_{\al-\ga}\,.
$$
We call these equations {\sl the modified hydrodynamics}
on the torus $T^2$.

\begin{rem}
Recall that the standard hydrodynamics of an ideal fluid on the torus $T^2$
is given by the Euler equation
$$
\partial_t\bfS=\{\bfS, \Delta^{-1}\bfS\}_1,
$$
on the vorticity function $\bfS$.
The definition of the Laplace operator uses metric, while to define the
modified hydrodynamics (\ref{cem})
we need to fix a complex structure on $T^2$.
\end{rem}

Now consider the Lax representation for the modified hydrodynamics.
In the double scaling limit we have
$$
\lim\vf_\ep(\al,z)=\exp(2\pi i \al_1z)\left(\f1{\al_\tau}+\f1{z}
\right)\,,
$$
$$
\lim f_\ep(\al,z)=-\exp(2\pi i \al_1z)\f1{\al_\tau^2}
$$
Therefore, the Lax pair (\ref{4.7}) in the rational limit is
$$
L(x_1,x_2;z)=-\sum_\al S_{\al}
\exp(2\pi i \al_1z)\left(\f1{\al_\tau}+\f1{z}\right)T_\al \,.
$$
$$
M(x_1,x_2)=-\sum_\al S_{\al} \exp(2\pi i \al_1z)\f1{\al_\tau^2}T_\al\,.
$$
We can drop the exponents in the both operators and
come to the following expressions
\beq{8.10}
L(x_1,x_2;z)=-\bp^{-1}\bfS(x_1,x_2)-\f1{z}\bfS(x_1,x_2)\,,
\eq
and
\beq{9.10}
M(x_1,x_2)=-\bp^{-2}\bfS(x_1,x_2)\,.
\eq
It is easy to see that the dispersionless Lax equation
$$
\p_t L(x_1,x_2;z)=\{L(x_1,x_2;z),M(x_1,x_2)\}
$$
is equivalent to the equations (\ref{cem}).

The infinite set of the integrals of motion comes from the decomposition
$$
(-1)^k\int_{T^2}L^k(x_1,x_2;z)dx_1dx_2=I_{0,k}+\sum_{s=2}^kI_{m,k}z^{-m}\,.
$$
The corresponding infinite hierarchy of the modified hydrodynamics
$$
\p_{s,t}\bfS=\{I_{s,k},\bfS\}_1
$$
has the dispersionless Lax form representation.

\medskip

{\sl 2. Bihamiltonian structure}\\
It turns out that the bihamiltonian structure survives in this limit. In fact,
 we have the following quadratic Poisson algebra $\cP_{0,\tau}$ on the
commutative torus $T^2$
\beq{6.10}
\{S_\al,S_0\}_2=
\sum_{\ga\neq\al}S_{\al-\ga}S_\ga(\f1{\ga_\tau^2}-
\f1{(\al_\tau-\ga_\tau)^2})
(\al\times\ga)\,,
\eq
\beq{7.10}
\{S_\al,S_\be\}_2=S_0S_{\al+\be}
(\al\times\be)+\sum_{\ga\neq\al,-\be}S_{\al-\ga}S_{\be+\ga}
f(\al,\be,\ga)(\ga\times(\al-\be))\,,
\eq
where
$$
f(\al,\be,\ga)=
\left(\f1{\ga_\tau}+\f1{\be_\tau-\al_\tau+\ga_\tau}
-\f1{\be_\tau+\ga_\tau}+\f1{\al_\tau-\ga_\tau}\right)\,,
~~\ga_\tau=\ga_1+\tau\ga_2\,.
$$
Similarly to the above,
we can consider  a one-parametric family of the quadratic algebras
by replacing $S_0\to S_0+\la$. Note  that it degenerates
to the standard Poisson brackets on $T^2$ in the limit $\la\to\infty$.
This allows us to define the bihamiltonian
structure for the  hierarchy of the modified hydrodynamics.
In particular, the equation (\ref{cem}) can be rewritten in the form as
$$
\p_t\bfS=\{S_0,\bfS\}_2\,.
$$
Such a bihamiltonian structure is  a curious feature in the modified
hydrodynamics, emphasizing its drastic difference from the classical
hydrodynamics.


\section{Quantum counterparts}
\setcounter{equation}{0}

In this section we present two associative algebras ``quantizing''
the discussed above two Poisson algebras in the case of the NCT.
It is easy with the linear bracket. Indeed, the quantization of
the Lie-Poisson algebra on $sin^*_\te$ leads to the universal
enveloping algebra $sin_\te$.

Replace the classical variables $\bfS$
on the NCT $\cA_\te$ by non-commuting variables
$\bhS$. Now  we consider the quantization of the
quadratic Poisson algebra $P_{\te,\ep,\tau}$ and
construct   the associative algebra $\hP_{\te,\ep,\tau,\hbar}$, where $\hbar\in E_\tau$
is the deformation parameter. For this we introduce the quantum
$R$-matrix related to the group $SIN_\te$.
The quantum Yang-Baxter equation defined on
$\GLN$ \cite{Be,Skl} is generalized to $SIN_\te$ in the following
way.
The quantum $R$-matrix on $SIN_\te\otimes SIN_\te$ assumes the form
$$
R(z,w)=\sum_c\vf_{\te,\ep}(c,z-w|\hbar)T_c\otimes T_{-c}\,,
$$
where $T_c$ is the basis of $SIN_\te$ and
the main ingredient in our construction is the function
$$
\vf_{\te,\ep}(c,z|\hbar)
=\bfe_\te(\ep_2c_2\te)\phi((\ep_1c_1+\tau \ep_2c_2)\te+\hbar,z)\,,~~\hbar\in
E_\tau\,.
$$
Note that, in contrast with the classical $r$-matrix (\ref{1.20}),
there is an additional term\\ $\vf_{\te,\ep}(0,z-w|\hbar)T_0\otimes T_0$.
\begin{predl}
The matrix $R$ satisfies the quantum
YB equation
\beq{qyb}
R_{12}(z-w)R_{13}(z)R_{23}(w)=R_{23}(w)R_{13}(z)R_{12}(z-w)\,.
\eq
\end{predl}
{\sl Proof.}\\
Consider the coefficients in front of $T_a\otimes T_b\otimes T_c$
in the l.h.s of (\ref{qyb}). It vanishes if $a+b+c\neq 0$.
Therefore (\ref{qyb}) implies
$$
\bfe_\te(a\times b+2b\times c)
\vf_{\te,\ep}(c,z-w|\hbar)\vf_{\te,\ep}(a-c,z|\hbar)\vf_{\te,\ep}(b+c,w|\hbar)
$$
$$
-\bfe_\te(a\times b+2b\times f)
\vf_{\te,\ep}(f,z-w|\hbar)\vf_{\te,\ep}(a-f,z|\hbar)\vf_{\te,\ep}(b+f,w|\hbar)
=0
$$
Here $a$, $b$ are fixed and $c$, $f$ are arbitrary elements of the lattice
$\mZ_{\te,\ep}(\tau)$
(\ref{Z}). Choose $f$ in the form $f=a-b-c$ to obtain
$$
\bfe_\te(a\times b+2b\times c)\left(
\vf_{\te,\ep}(c,z-w|\hbar)\vf_{\te,\ep}(a-c,z|\hbar)\vf_{\te,\ep}(b+c,w|\hbar)
\right.
$$
$$
\left.
-\vf_{\te,\ep}(a-b-c,z-w|\hbar)
\vf_{\te,\ep}(a-c,w|\hbar)\vf_{\te,\ep}(b+c,z|\hbar)
\right)
=0\,.
$$
This equality can be transform to the form
$$
\vf_{\te,\ep}(a+2\hbar,z|\hbar)\vf_{\te,\ep}(b,w|\hbar)
\bfe_\te(a\times b+2b\times c)
$$
$$
\times\left(\ze_{\te,\ep}(c+\hbar)
-\ze_{\te,\ep}(a-b-c+\hbar)+\ze_{\te,\ep}(a-c+\hbar)-\ze_{\te,\ep}(b+c+\hbar)
\right)=0\,
$$
by means of (\ref{ir}).
If $b=0$ then the l.h.s. vanishes and we come to (\ref{qyb}).
Let $b\neq 0$ and consider the shift
 $c\to c+jb$, where $j\in\mZ$. The shift does not change the exponential
factor.
Take the sum over the orbit generated by the shifts
$$
\sum_{j\in\mZ}
\ze_{\te,\ep}(c+jb+\hbar)-\ze_{\te,\ep}(a-b-jb-c+\hbar)
+\ze_{\te,\ep}(a-c+jb+\hbar)-\ze_{\te,\ep}(b+c+jb+\hbar)\,.
$$
One can see that the neighboring terms in the series vanish
and we come to (\ref{qyb}) for an arbitrary $b$.
$\Box$

This Proposition allows us to define
the associative algebra $\hP_{\te,\ep,\tau,\hbar}$  by the relation
\beq{1.9}
R(z-w)L^\hbar_1(z)L^\hbar_2(w)=L^\hbar_2(w)L^\hbar_1(z)R(z-w)\,,
\eq
where
$$
L^\hbar(z)=\hS_0\vf_{\te\ep}(0,z|\hbar)T_0
+\sum_\al \hS_{-\al}\vf_{\te,\ep}(\al,z|\hbar)T_\al\,.
$$
In order to obtain the relations in $\hP_{\te,\ep,\tau,\hbar}$
one should exclude the spectral parameters $z,w$ from (\ref{1.9}).

\begin{lem}
The relations in the associative algebra $\hP_{\te,\tau,\hbar}$
 assume the form
\beq{2.9}
\sum_c\hS_{a-c}\hS_{b+c}\bfe_\te(c\cdot(a-b))\bff_{\te,\ep}(a,b,c|\hbar)=0\,,
~~{\rm for~any~}a,b\in\mZ\oplus\mZ\,,
\eq
where
\beq{3.9}
\bff_{\te,\ep}(a,b,c|\hbar)=
\ze_{\te,\ep}(c+\hbar)-\ze_{\te,\ep}(a-b-c+\hbar)-\ze_{\te,\ep}(b+c+\hbar)+
\ze_{\te,\ep}(a-c+\hbar)
\eq
$$
=-\frac{
\vth'(0)\vth_{\te,\ep}(\al+2\hbar)\vth_{\te,\ep}(\be)\vth_{\te,\ep}(\be-\al+2\ga)
}{
\vth_{\te,\ep}(\al-\ga+\hbar)\vth_{\te,\ep}(\be+\ga+\hbar)
\vth_{\te,\ep}(\be-\al+\ga+\hbar)\vth_{\te,\ep}(\ga+\hbar)
}\,.
$$
and $\vth_{\te,\ep},~\ze_{\te,\ep}(.)$ are the constants (\ref{C0}), (\ref{C1}).
\end{lem}
{\sl Proof.}\\
Consider in (\ref{1.9}) the matrix element $T_a\otimes T_b$.
We come to the relation
\beq{4.9e}
\sum_c\hS_{a-c}\hS_{b+c}\{
\vf_{\te,\ep}(c,z-w|\hbar)\vf_{\te,\ep}(a-c,z|\hbar)\vf_{\te,\ep}(b+c,w|\hbar)
\eq
$$
-\vf_{\te,\ep}(a-b-c,z-w|\hbar)\vf_{\te,\ep}(a-c,w|\hbar)\vf_\te(b+c,z|\hbar)
\}\bfe_\te(c\cdot(a-b))=0\,.
$$
The expression in the brackets $\{~\}$ is equal to
$$
\bfe_\te((a+2\hbar)z+(b+\hbar)w)
\{\phi_{\te,\ep}(c,z-w|\hbar)\phi_{\te,\ep}(a-c,z|\hbar)
\phi_{\te,\ep}(b+c,z-w|\hbar)
$$
$$
-\phi_{\te,\ep}(a-b-c,z-w|\hbar)\phi_{\te,\ep}(a-c,w|\hbar)
\phi_{\te,\ep}(b+c,z|\hbar)
\}\,,
$$
where $\phi_{\te,\ep}(c,z|\hbar)=\phi((\ep_1c_1+\tau \ep_2c_2)\te+\hbar,z)$.
Finally, by using (\ref{ir}) we come from (\ref{4.9e}) to (\ref{3.9}). $\Box$
\bigskip

We rewrite (\ref{2.9}) in the form most resembling the
original Sklyanin relations \cite{Skl}. Let ${\bf\gA}(\al,\be)$
(respectively, ${\bf \gS}(\al,\be))$ be the operator of
antisymmetrization (respectively, symmetrization) with respect
to the permutations of two indices  $(\al,\be)$.

\begin{predl}
The relations (\ref{3.9}) are equivalent to the commutator relations
\beq{1.q}
[\hS_0,\hS_b]=
-\sum_{\ga\neq b}\hS_{-\ga}\hS_{b+\ga}\bfe_\te(b\cdot\ga)
\frac{\bff_{\te,\ep}(0,b,\ga|\hbar)}
{\bff_{\te,\ep}(0,b,0|\hbar)}\,,
\eq
\beq{2.q}
[\hS_\al,\hS_\be]={\bf\gA} (\al,\be)
\left\{
\left(
\hS_{\al+\be}\hS_0\bfe_\te(-\be\cdot\al)-\hS_0\hS_{\al+\be}\bfe_{\te,\ep}(\be\cdot\al)
\right)\frac{
\bff_{\te,\ep}(\al,\be,\al|\hbar)}
{\bff_{\te,\ep}(\al,\be,0|\hbar)}\right.
\eq
$$
\left.
+\sum_{\ga\neq\al,-\be}\left(
\hS_{\al-\ga}\hS_{\be+\ga}\bfe_\te(\ga\cdot(\al-\be))-
\hS_{\be+\ga}\hS_{\al-\ga}\bfe_\te(-\ga\cdot(\al-\be))
\right)\frac{
\bff_{\te,\ep}(\al,\be,\ga|\hbar)}
{\bff_{\te,\ep}(\al,\be,0|\hbar)}\right\}
\,,
$$
where $\bfC_\te(\al,\be)$ are $\sln$ structure constants (\ref{AA4}), and
$$
{\bf \gS}(\al,\be)
\left\{
\left(
\hS_{\al+\be}\hS_0\bfe_\te(-\be\cdot\al)-\hS_0\hS_{\al+\be}\bfe_{\te,\ep}(\be\cdot\al)
\right)
\frac{\bff_{\te,\ep}(\al,\be,\al|\hbar)}
{\bff_{\te,\ep}(\al,\be,0|\hbar)}\right.
$$
$$
\left.
+\sum_{\ga\neq\al,-\be}\left(
\hS_{\al-\ga}\hS_{\be+\ga}\bfe_\te(\ga\cdot(\al-\be))-
\hS_{\be+\ga}\hS_{\al-\ga}\bfe_\te(-\ga\cdot(\al-\be))
\right)
\frac{
\bff_{\te,\ep}(\al,\be,\ga|\hbar)}
{\bff_{\te,\ep}(\al,\be,0|\hbar)}\right\}
=0\,.
$$
\end{predl}
{\sl Proof.}\\
The relation (\ref{2.9}) can be rewritten in  the form
$$
\sum_c\left(
\hS_{a-c}\hS_{b+c}\bfe_\te(c\cdot(a-b))-\hS_{b+c}\hS_{a-c}\bfe_\te(c\cdot(a-b))
\right)
\bff_{\te,\ep}(a,b,c|\hbar)=0\,.
$$
due to the equality
$\bff_{\te,\ep}(a,b,c|\hbar)=-\bff_{\te,\ep}(a,b,a-b-c|\hbar)$.
For the case $a=0$ this relation assumes the form (\ref{1.q}).
To come to (\ref{2.q}) we  single out the
two terms with $c=0$ and put them in the left hand side.
In the right hand side we first
write down the terms with $c=a$ and $c=-b$ and obtain the required relations.
$\Box$

\smallskip
It would be interesting to describe a quantum version of the whole
bihamiltonian structure for elliptic rotators
on $sin^*$ or $Ham(T^2)$. Posing the problem more generally,
the bihamiltonian recursion precedure
of generating the conserved quantities from a Casimir function for a
linear family of Poisson structures (see \cite{Mag}) might have
a quantum analog as an expansion of a central element for a linear
family of associative algebras. The latter seems to be a very strong
requirement on a pair of associative algebras
and it would be very interesting to find any non-trivial example
of this kind. Namely, one is looking for a pair of associative algebras,
such that their mixed associator satisfies some consistency condition.
 This would provide the most straightforward quantization
for a system with a bihamiltonian structure.


\section*{Acknowledgments}
\addcontentsline{toc}{section}{\numberline{}Acknowledgments}
We are grateful for hospitality to the Max-Planck-Institut fur Mathematik
in Bonn, where most of this work was done. B.K. is also grateful 
for hospitality to the University of Nice and  IHES in Bures-sur-Yvette. 
The work of B.K. was  partially supported by an NSERC research grant.
The work of A.L. and M.O. was supported in part by grants RFBR-03-02-17554,
NSch-1999-2003.2 and CRDF  12729.


\section{Appendix}

\subsection{Appendix A. Elliptic functions.}
\setcounter{equation}{0}
\def\theequation{A.\arabic{equation}}

Here we summarize the main formulae for elliptic functions.
Consider an elliptic curve
\beq{bsc}
E_\tau=\mC/(\mZ+\tau\mZ)\,,~~
q=\bfe(\tau)=\exp (2\pi i\tau)\,.
\eq

The basic element in our consideration is the theta  function:
\beq{A.1a}
\vth(z;\tau)=q^{\frac
{1}{8}}\sum_{n\in {\bf Z}}(-1)^ne^{\pi i(n(n+1)\tau+2nz)}=
\eq
$$
q^{\frac{1}{8}}e^{-\frac{i\pi}{4}} (e^{i\pi z}-e^{-i\pi z})
\prod_{n=1}^\infty(1-q^n)(1-q^ne^{2i\pi z})(1-q^ne^{-2i\pi z})\,.
 $$
\bigskip

{\sl 1. The Weierstrass functions:}
\beq{A0}
\sigma(z;\tau)=\exp(\eta z^2)\frac{\vth(z;\tau)}{ \vth'(0;\tau)}\,,
\eq
where
\beq{A00}
\eta(\tau)=-\f1{6}\frac{\vth'''(0;\tau)}{ \vth'(0;\tau)}\,.
\eq
\beq{A.1}
\ze(z;\tau)=\p_z\log\vth(z;\tau)+2\eta(\tau)z\,, ~~
\ze(z;\tau) \sim\f1{z}+O(z^3)\,.
\eq

\beq{A.2}
\wp(z;\tau)=-\p_z\ze(z;\tau)\,.
\eq

\beq{4.0}
\wp(u;\tau)=\f1{u^2}+
+\sum'_{j,k}\left(
\f1{(j+k\tau+u)^2}-\f1{(j+k\tau)^2}\right)\,.
\eq

\bigskip
\noindent
{\sl 2. Function $\phi$:}
\beq{A.3}
\phi(u,z)=
\frac
{\vth(u+z)\vth'(0)}
{\vth(u)\vth(z)}
\eq
It has a pole at $z=0$ and the expansion
\beq{A.3a}
\phi(u,z)=\frac{1}{z}+\zeta(u|\tau)+2\eta(\tau)z+
\frac{z}{2}((\zeta(u;\tau)+2\eta(\tau)z)^2-\wp(u))+\ldots\,,
\eq
as $z\to 0$. Furthermore,
\beq{A3b}
\phi(u,z)^{-1}\p_u\phi(u,z)=\zeta(u+z)-\zeta(u)+2\eta(\tau)z\,.
\eq
\beq{A.7}
\phi(u,z)=\exp(-2\eta uz)
\frac
{\si(u+z)}{\si(u)\si(z)}\,,
\eq
\beq{A.7a}
\phi(u,z)\phi(-u,z)=\wp(z)-\wp(u)\,.
\eq

\bigskip
\noindent
{\sl 3. Quasi-periodicity:}
\beq{aa1}
\vth(z+1;\tau)=-\vth(z;\tau)\,,~~\vth(z+\tau;\tau)=\
-\bfe(-\oh\tau-z)\vth(z;\tau)\,.
\eq
\beq{aa2}
\ze(z+1;\tau)=\ze(z;\tau)-2\eta\,,~~
\ze(z+\tau;\tau)=\ze(z;\tau)-2(\pi i +\eta\tau)\,.
\eq
\beq{aa3}
\wp(u+1;\tau)=\wp(u+\tau;\tau)=\wp(u;\tau)\,.
\eq
\beq{aa4}
\phi(u+1,z)=\phi(u,z)\,,~~\phi(u+\tau,z)=\bfe(-z)\phi(u,z)\,.
\eq

\bigskip
\noindent
{\sl 4. Parity:}
\beq{aa5}
\vth(-z;\tau)=-\vth(z;\tau)\,.
\eq
\beq{aa6}
\ze(-z;\tau)=-\ze(z;\tau)\,.
\eq
\beq{aa7}
\wp(-u;\tau)=\wp(u;\tau)\,.
\eq
\beq{aa8}
\phi(-u,-z)=-\phi(u,z)\,.
\eq

\bigskip
\noindent
{\sl 5. The Fay three-section formula:}

\beq{ad3}
\phi(u_1,z_1)\phi(u_2,z_2)-\phi(u_1+u_2,z_1)\phi(u_2,z_2-z_1)-
\phi(u_1+u_2,z_2)\phi(u_1,z_1-z_2)=0\,.
\eq
Particular cases of this formula are (\ref{A.7a}),
the  Calogero functional equation
\beq{ad2}
\phi(u,z)\p_v\phi(v,z)-\phi(v,z)\p_u\phi(u,z)=(\wp(v)-\wp(u))\phi(u+v,z)\,,
\eq
 and
\beq{ad4}
\phi(u_1,z)\phi(u_2,z)-\phi(u_1+u_2,z)
(\zeta(u_1)+\zeta(u_2)-2\eta(\tau)(u_1+u_2))+
\p_z\phi(u_1+u_2,z)=0\,.
\eq
For $u_1+u_2+u_3=0$,
\beq{4.15}
\phi(u_1,z)\phi(u_2,z)\phi(u_3,z)=
\left[\wp(z)-\wp(u_3)\right]
\left[\ze(u_1)+ \ze(u_2)+ \ze(u_3-z)+ \ze(z)\right]\,,
\eq
as follows from (\ref{A3b}), (\ref{A.7a}),and (\ref{ad4}).
Then
\beq{ad5}
\phi(u_1,z)\phi(u_2,z)\phi(u_3,z)|_{z\to 0}=
 \f1{z^3}
+\f1{z^2}\left[\ze(u_1)+ \ze(u_2)+ \ze(u_3)\right]
\eq
$$
-\oh\wp'(u_3)-\wp(u_3)
\left[\ze(u_1)+ \ze(u_2)+ \ze(u_3)\right] +O(z)\,.
 $$
>From (\ref{ad3}) and (\ref{A3b}) we have
\beq{A20}
\phi(u_1,z)\phi(u-u_1,z)=\phi(u,z)(\ze(u_1)+\ze(u-u_1)-\ze(u+z)+
\ze(z))\,.
\eq
Another important relation is
\beq{ir}
\phi(v,z-w)\phi(u_1-v,z)\phi(u_2+v,w)
-\phi(u_1-u_2-v,z-w)\phi(u_2+v,z)\phi(u_1-v,w)=
\eq
$$
\phi(u_1,z)\phi(u_2,w)f(u_1,u_2,v)\,,
$$
where
\beq{ir1}
f(u_1,u_2,v)=\ze(v)-\ze(u_1-u_2-v)+\ze(u_1-v)-\ze(u_2+v)\,.
\eq
One can rewrite the last function as
\beq{ir3}
f(u_1,u_2,v)=-\frac{
\vth'(0)\vth(u_1)\vth(u_2)\vth(u_2-u_1+2v)
}{
\vth(u_1-v)\vth(u_2+v)\vth(u_2-u_1+v)\vth(v)
}\,.
\eq
To prove (\ref{ir}) one should consider the expression
$$
\frac{
\phi(v,z-w)\phi(u_1-v,z)\phi(u_2+v,w)
-\phi(u_1-u_2-v,z-w)\phi(u_2+v,z)\phi(u_1-v,w)
}
{\phi(u_1,z)\phi(u_2,w)}\,.
$$
It is a doubly periodic function
in $z$ and $w$ without poles. Therefore, it
is a constant, which depends on $u_1,u_2,v$. This constant is equal to
$f(u_1,u_2,v)$ (\ref{ir1}).
It is easy to check that the elliptic functions (\ref{ir1}) and (\ref{ir3})
coincide.

A particular case of (\ref{ir}), which corresponds to the case
$v=u_1$ (or $v=-u_2$), is the Fay identity (\ref{ad3}). Another
particular case comes from
$u_2=0$ (or $u_1=0$):
\beq{ir2}
\phi(v,z-w)\phi(u-v,z)\phi(v,w)-\phi(u-v,z-w)\phi(v,z)\phi(u-v,w)=
\eq
$$
\phi(u_1,z)(\wp(v)-\wp(u-v))\,.
$$

\subsection{Appendix B. Elliptic constants related
to $\gln$.}
\setcounter{equation}{0}
\def\theequation{B.\arabic{equation}}

Let $1\leq k<N$ be a coprime number with respect to $N$ and set
$\te=\frac{k}{N}$. Define
\beq{AA0}
{\bf e}_\te(z)=\exp (2\pi i\te z)\,,
\eq
\beq{AA11}
Q=\di({\bf e}_\te(1),\ldots,{\bf e}_\te(m),\ldots,1)\,
\eq
\beq{AA2}
\La=
\left(\begin{array}{ccccc}
0&1&0&\cdots&0\\
0&0&1&\cdots&0\\
\vdots&\vdots&\ddots&\ddots&\vdots\\
0&0&0&\cdots&1\\
1&0&0&\cdots&0
\end{array}\right)\,.
\eq

Consider a finite two-dimensional lattice of order $N^2$
$$
 \mZ^{(2)}_N=\mZ/N\mZ\oplus\mZ/N\mZ\,.
$$
The matrices
\beq{AA3}
T_{a}=\f1{2\pi i\te}\bfe_\te(\frac{a_1a_2}{2})Q^{a_1}\La^{a_2}\,,
~(a=(a_1,a_2)\in\mZ^{(2)}
\eq
generate the basis in $\gln$.
We use the Greek letters for the elements of the lattice
\beq{lat}
\ti{\mZ}^{(2)}_N=\mZ^{(2)}_N\setminus(0,0)\,.
\eq
 Thus, $\{T_\al\}$ define a basis in $\sln$.
Since
\beq{AA3a}
T_aT_b=\f1{2\pi\te}\bfe_\te(-\frac{a\times b}{2})T_{a+b}\,, ~~
(a\times b=a_1b_2-a_2b_1)
\eq
 the commutation relations in this basis assume the form
\beq{AA4b}
[T_{\al},T_{\be}]=\bfC_\te(\al,\be)T_{\al+\be}\,,~~
\eq
where
\beq{AA4}
\bfC_\te(\al,\be)=\f1{\pi\te}\sin\pi\te(\al\times \be)\,.
\eq

Let
\beq{AA4a}
\mZ_\te^{(2)}(\tau)=(\ga_1+\ga_2\tau)\te\,,~~\ga\in\ti{\mZ}^{(2)}_N
\eq
be a regular lattice of order $N^2-1$ on $E_\tau$.
Introduce the following  constants on $\mZ_\te^{(2)}(\tau)$:
\beq{AA50}
\vth_\te(\ga)=\vth((\ga_1+\ga_2\tau)\te)\,,
\eq
\beq{AA5}
\ze_\te(\ga)=\ze((\ga_1+\ga_2\tau)\te)\,,
~~\wp_\te(\ga)=\wp((\ga_1+\ga_2\tau)\te)\,,
\eq
and the quasi-periodic functions on $E_\tau$
\beq{phi}
\phi_\te(\ga,z)=\phi((\ga_1+\ga_2\tau)\te,z)\,,
\eq
\beq{vf}
\vf_\te(\ga,z)=\bfe_\te(\ga_2z)\phi_\te(\ga,z)\,,
\eq
\beq{f}
f_\te(\ga,z)=
\bfe_\te(\ga_2z)\p_u\phi(u,z)|_{u=(\ga_1+\ga_2\tau)\te}\,.
\eq
It follows from (\ref{A.3}) that
\beq{qp}
\vf_\te(\ga,z+1)=\bfe_\te(\ga_2)\vf_\te(\ga,z)\,,~~
\vf_\te(\ga,z+\tau)\bfe_\te(-\ga_1)\vf_\te(\ga,z)\,.
\eq

\subsection{Appendix C. Non-commutative torus}
\setcounter{equation}{0}
\def\theequation{C.\arabic{equation}}

\noindent
{\sl 1. Definition and representation.}\\
The non-commutative torus ${\cal A}_\te$ is a unital algebra with the
 two generators $(U_1,U_2)$ that satisfy the relation
\beq{3.1}
U_1U_2=\bfe_\te^{-1} U_2U_1,~\bfe_\te=e^{ 2\pi i \te},~ \te\in[0,1)\,.
\eq

Elements of ${\cal A}_\te$ are the double sums
$$
{\cal A}_\te=
\left\{X=\sum_{a_1,a_2\in{\mathbb Z}}c_{a_1,a_2}U_1^{a_1}U_2^{a_2}~|~~
c_{a_1,a_2}\in\mathbb C\right\}\,,
$$
where $c_{a_1,a_2}$ are elements of the ring ${\gS}$
of  rapidly decreasing sequences on ${\mathbb Z}^2$
\beq{3.1a}
\gS=\{c_{a_1,a_2}~|~~
{\rm sup}_{a_1,a_2\in\mZ}(1+a_1^2+a_2^2)^k|c_{a_1,a_2}|^2<\infty~~
{\rm for all }~k\in \mN\}\,.
\eq
The trace functional $\tr(x)$ on ${\cal A}_\te$ is defined by
\beq{tr}
\tr(X)=c_{00}\,.
\eq
It  satisfies the evident identities
$$
\tr(1)=1,~~\tr(XY)=\tr(YX)\,.
$$
The dual space to $\gS$ is the space
\beq{ds}
\gS'=\{S_{a_1,a_2}~|~\sum_{a_1,a_2} c_{a_1,a_2}S_{-a_1,-a_2}<\infty,
~ c_{a_1,a_2}\in\gS\}\,.
\eq

The relation with the commutative algebra of smooth functions
on the two-dimensional torus
\beq{T}
T^2=\{\mR^2/\mZ\oplus\mZ\}\,\sim \,\{0<x_1\leq 1,\,0<x_2\leq 1\}.
\eq
comes from the identification
\beq{3.2}
U_1\to\bfe(x),~~U_2\to\bfe(y),
\eq
while the multiplication on $T^2$ becomes the Moyal multiplication:
\beq{3.3}
(f*g)(x):=fg+
\sum_{n=1}^\infty\frac{(\pi\te)^n}{n!}
\ve_{r_1s_1}\ldots\ve_{r_ns_n}(\p^n_{r_1\ldots r_n}f)
(\p^n_{s_1\ldots s_n}g).
\eq
The trace functional (\ref{tr}) in the Moyal identification
is the integral
\beq{3.6}
\tr f=-\f1{4\pi^2}\int_{\cA_\te}fdx_1dx_2=f_{00}\,.
\eq

 We can identify $U_1,U_2$ with matrices from GL$(\infty)$.
Define GL$(\infty)$ as the associative algebra of infinite matrices
$c_{jk}E_{jk}$, where $E_{jk}=||\delta_{jk}||$, such
that
$$
{\rm sup}_{j,k\in\mZ}|c_{jk}|^2|j-k|^n<\infty\, ~{\rm for}~n\in\mN\,.
$$
Consider the following two matrices from GL$(\infty)$:
$$
Q=\di (\bfe(j\te))~~{\rm and }~~\La=||\delta_{j,j+1}||\,,~j\in\mZ\,.
$$
We have the following identification
\beq{3.5}
U_1\to Q,~U_2\to \La\,.
\eq

Another useful realization of $\cA_\te$ in the Schwartz space on $\mR$
 by the operators
\beq{3.5a}
U_1f(x)=f(x-\te)\,,~~U_2f(x)=\exp(2\pi ix)f(x)\,.
\eq

\bigskip
\noindent
{\sl 2. $sin$-algebra.}\\
Define the following quadratic combinations of the generators
\beq{3.10}
T_\al=\frac{i}{2\pi\te}\bfe
\left(\frac{\al_1\al_2}{2}\te
\right)U_1^mU_2^n\,~~\al\in\ti{\mZ}^{(2)}\,,
\eq
$$
\ti{\mZ}^{(2)}=\{\al=(\al_1,\al_2)\,,\al_j\in\mZ,\,\al\neq(0,0)\}\,.
$$
Their commutator has the form of the sin-algebra
\beq{3.11}
[T_{\al},T_{\be}]=\bfC_\te(\al,\be)T_{\al+\be}\,,
\eq
where
\beq{3.11a}
\bfC_\te(\al,\be)=\f1{\pi\te}\sin\pi\te(\al\times \be)\,.
\eq
We denote by $sin_\te$ the Lie algebra with the generators $T_{\al}$
over the ring $\gS$ (\ref{3.1a})
\beq{3.12}
\psi=\sum_{\al}\psi_{\al}T_{\al},~~\psi_{\al}\in\gS\,,
\eq
 and by $SIN_\te$ the group of invertible elements
from ${\cal A}_\te$. The coalgebra $sin^*_\te$ is the linear space
$$
sin^*_\te=\left\{
\bfS=\sum_{\al\in\ti{\mZ}^{(2)}}s_{\al}T{\al},~s_{\al}\in\gS'\right\}\,.
$$

In the Moyal representation (\ref{3.3}) the commutator of $sin_\te$
has the form
\beq{3.14}
[f(x,y),g(x,y)]=\{f,g\}^\te:=\f1{\te}(f*g-g*f)\,
\eq

\bigskip
\noindent
{\sl 3. Elliptic constants related to NCT $\cA_\te$}\\
  Introduce two  numbers
 $\ep=(\ep_1,\ep_2)$   such that  $\ep_a\te<1$ and $\ep_a\te$
are irrational. Consider the  dense
set  $\mZ_{\te,\ep}(\tau)$ in $E_\tau$:
\beq{Z}
\mZ_{\te,\ep}(\tau)=\{(\ep_1\ga_1+\tau\ep_2\ga_2)\te\in E_\tau~~|~
~(\ga_1,\ga_2)\in\ti{\mZ}^{(2)}\}\,.
\eq
The corresponding
elliptic functions with the arguments from $\mZ_{\te,\ep}(\tau)$
 are as follows:
\beq{C0}
\vth_{\te,\ep}(\ga)=\vth((\ep_1\ga_1+\ep_2\ga_2\tau)\te)\,,
\eq
\beq{C1}
\ze_{\te,\ep}(\ga)=\ze((\ep_1\ga_1+\tau\ep_2\ga_2)\te)\,,
~~
\wp_{\te,\ep}(\ga)=\wp((\ep_1\ga_1+\tau\ep_2\ga_2)\te)\,.
\eq
\beq{CC2}
\phi_{\te,\ep}(\ga,z)=\phi(-(\ep_1\ga_1+\tau\ep_2\ga_2)\te,z)\,,
\eq
\beq{CC3}
\vf_{\te,\ep}(\ga)=\bfe_\te(\ep_2\ga_2z)\phi_{\te,\ep}(\ga,z)\,.
\eq
\beq{CC3a}
f_{\te,\ep}(\ga)=\bfe_\te(\ep_2\ga_2z)\p_u\phi(u,z)|_{u=(\ep_1\ga_1+\tau\ep_2\ga_2)\te}\,.
\eq
\
\bigskip
\noindent
{\sl 4. Dispersionless limit.}\\
In the limit $\te\to 0$ the Lie algebra $sin_\te$ becomes
the Lie algebra of Hamiltonian functions
\beq{Ham}
Ham(T^2)\sim C^\infty(T^2)/\mC
\eq
 equipped with the canonical Poisson brackets.
In $Ham(T^2)$
 we have the Fourier
basis
\beq{fb}
\bfe(\al\cdot x)=\exp (2\pi i(\al_1x_1+\al_2x_2))
\eq
instead of the basis (\ref{3.10}).
 The commutator
(\ref{3.11}) becomes
$$
[\bfe(\al x),\bfe(\be x)]=(\al\times\be)\bfe((\al+\be)\cdot x)\,.
$$
The algebra $Ham(T^2)$ (without constant Hamiltonians) is isomorphic
to Lie algebra $SVect_0(T^2)$
of the divergence-free {\it zero-flux} vector fields on $T^2$
equipped with the area
form $-4\pi^2dx_1dx_2$. Let $h(x_1,x_2)\in Ham(T^2)$. Then
the Hamiltonian field $V_h$ corresponding to the Hamiltonian function $h$ is
\beq{3.20}
V_h=-\f1{4\pi^2}((\p_2h)\p_1-(\p_1h)\p_2)\,,
\eq
while
\beq{3.21}
[V_{h},V_{h'}]=V_{\{h,h'\}}\,.
\eq
For $f(x)=\sum_\al f_\al\bfe(\al\cdot x)$
\beq{int}
\int_{T^2}f=-\f1{4\pi^2}f_0\,.
\eq


\end{document}